\documentclass[aps, pra, reprint, superscriptaddress, floatfix, longbibliography]{revtex4-2}
\usepackage[utf8]{inputenc}
\usepackage{siunitx}
\usepackage{amsmath,amssymb}
\usepackage{xcolor}
\usepackage[colorlinks=true]{hyperref}
\usepackage{graphicx}
\usepackage{multirow}
\usepackage{microtype}
\usepackage{soul}

\graphicspath{{FIGURES/}}

\usepackage[draft]{changes}
\newcommand{\ket}[1]{\ensuremath{|#1\rangle}}
\newcommand{\bra}[1]{\ensuremath{\langle #1|}}

\begin{document}

\title{Strategies and trade-offs for controllability and memory time of ultra-high-quality microwave cavities in circuit QED}

\author{Iivari Pietik\"{a}inen}
\thanks{These authors contributed equally}
\affiliation{Department of Optics, Palack\'{y} University, 17. listopadu 1192/12, 77900 Olomouc, Czechia}

\author{Ond\v{r}ej \v{C}ernot\'{i}k}
\thanks{These authors contributed equally}
\affiliation{Department of Optics, Palack\'{y} University, 17. listopadu 1192/12, 77900 Olomouc, Czechia}

\author{Alec Eickbusch}
\thanks{Present address: Google Quantum AI, Santa Barbara, CA}
\affiliation{Yale Quantum Institute, PO Box 208 334, 17 Hillhouse Ave, New Haven, CT 06520-8263, USA}
\affiliation{Departments of Applied Physics and Physics, Yale University, New Haven, CT 06511, USA}

\author{Aniket Maiti}
\affiliation{Yale Quantum Institute, PO Box 208 334, 17 Hillhouse Ave, New Haven, CT 06520-8263, USA}
\affiliation{Departments of Applied Physics and Physics, Yale University, New Haven, CT 06511, USA}

\author{John W. O. Garmon}
\affiliation{Yale Quantum Institute, PO Box 208 334, 17 Hillhouse Ave, New Haven, CT 06520-8263, USA}
\affiliation{Departments of Applied Physics and Physics, Yale University, New Haven, CT 06511, USA}

\author{Radim Filip}
\affiliation{Department of Optics, Palack\'{y} University, 17. listopadu 1192/12, 77900 Olomouc, Czechia}

\author{Steven M. Girvin}
\affiliation{Yale Quantum Institute, PO Box 208 334, 17 Hillhouse Ave, New Haven, CT 06520-8263, USA}
\affiliation{Departments of Applied Physics and Physics, Yale University, New Haven, CT 06511, USA}

\date{\today}

\begin{abstract}
    Three-dimensional microwave cavity resonators have been shown to reach lifetimes of the order of a second by maximizing the cavity volume relative to its surface, using better materials, and improving surface treatments. Such cavities represent an ideal platform for quantum computing with bosonic qubits, but their efficient control remains an outstanding problem since the large mode volume results in inefficient coupling to nonlinear elements used for their control. Moreover, this coupling induces additional cavity decay via the inverse Purcell effect which can easily destroy the advantage of {a} long intrinsic lifetime. Here, we discuss conditions on, and protocols for, efficient utilization of these ultra-high-quality microwave cavities as memories for conventional superconducting qubits. We show that, surprisingly, efficient write and read operations with ultra-high-quality cavities does not require similar quality factors for the qubits and other nonlinear elements used to control them. Through a combination of analytical and numerical calculations, we demonstrate that efficient coupling to cavities with second-scale lifetime is possible with state-of-the-art transmon and SNAIL devices and outline a route towards controlling cavities with even higher quality factors. Our work explores a potentially viable roadmap towards using ultra-high-quality microwave cavity resonators for storing and processing information encoded in bosonic qubits. 
\end{abstract}

\maketitle

\section{Introduction}

High-quality (high-$Q$) superconducting microwave resonators \cite{Haroche_RevModPhys.73.565,Haroche_RevModPhys.85.1083,Brecht2015c,Brecht2016, Reagor2016,Romanenko_PhysRevApplied.13.034032,BlaiscQEDReviewRMP2020,SchusterCavity2022BAPS,RosenblumCavity2023,krasnok2023advancements,zhao2023integrating} 
can be used to store continuous-variable quantum information as well as error-correctable logical qubits based on bosonic encodings. They offer the twin benefits of a large Hilbert space within a single quantum mode inside a simple piece of hardware. Furthermore, they have a simple error model (amplitude damping with little intrinsic dephasing \cite{Sivak_GKP_2022,Rosenblum2018,Eickbusch2021X}) which offers strong advantages for quantum error correction \cite{Girvin_LesHouches_QEC} relative to traditional logical qubits based on collections of physical qubits. Indeed bosonic codes \cite{GKP2001,BinomialCodes,GrimsmoPuri_GKP_2021,Ofek2016,Hu2019,LuyanSun2020,LuyanSun2022,ChenWangAutonomous,
HOME-GKP2019,RoyerGKP1_2020,Campagne2020,HomeGKPQEC2022,Sivak_GKP_2022} 
are, to date, the only architecture that has exceeded the break-even point for quantum error correction of a memory, where the error-corrected logical qubit maintains coherence longer than any physical qubit in the system. 

Classical computer architectures contain highly heterogeneous hardware elements including specialized fast processor memory, cache memory and longer-term memory elements with slow read-write access. Such architectures also require the means to move data seamlessly across different memory types. We may reasonably require a similar memory hierarchy for quantum computer systems \cite{Thaker_Q_Memory_Hierarchies_10.1109/ISCA.2006.32,liu2023quantum_memory,QASMBENCH_10.1145/3550488,stein2023microarchitectures}.

In executing quantum algorithms we can expect that there {may be} circumstances in which it would be valuable to have memory elements optimized for longer-term storage rather than fast access. Here, the figure of merit might simply be wall-clock storage time (assuming at least moderately good read-write fidelity). Other times, it might be beneficial to have long-lived elements (similar to classical cache memory) that act as logical qubits for computational purposes, requiring fast, high-fidelity gates. Here, the figure of merit would be the one- and two-logical qubit gate fidelities which would (at a minimum) require a high ratio of memory time to gate operation time as one underlying figure of merit.

In addition, algorithms may require some qubits to be entangled with others but then remain idle for many clock cycles before the next gate is applied to them. In such a case, it would be advantageous to temporarily move the quantum information from such a \emph{compute qubit} to a \emph{memory qubit}. An important example of this situation is the quantum Fourier transform (QFT) algorithm. The QFT and various other \emph{proxy apps} have been {cataloged} in {Ref.}~\cite{QASMBENCH_10.1145/3550488} along with metrics including \emph{gate density} and \emph{retention lifespan} that are directly related to the memory/compute ratio discussed here. Finally, complex error-correcting codes capable of long storage times could also result in longer gate times. Depending on specifics, these might follow from the relaxed requirements on control speed (thanks to the longer storage times) or stem from complicated gate sequences needed to control multi-photon states (leading to long gate times even with fast control). All of these factors need to be considered when exploring the design space for both individual hardware elements and system architectures.

With discrete-variable superconducting qubits, the two quadratures of a classical microwave drive can be used to produce Rabi rotations around the $x$ and $y$ axes of the Bloch sphere and thus suffice for universal single-qubit control. In contrast, superconducting resonators are harmonic oscillators with evenly spaced energy levels. As a result, classical microwave drives that couple linearly to the oscillator position and momentum produce only coherent states (classical displacements of the quantum vacuum state) and are not sufficient for the universal control required to produce photon number Fock states, bosonic code states, and other more general non-Gaussian operations such as quantum error correction~\cite{Grimsmo2020,Ma2021,Brady2024}. While {simple} phase-space displacements do not require auxiliary qubits, Gaussian single- and two-mode squeezing operations \cite{LehnertSqueezingPhysRevLett.106.220502,HAYSTAC_Backes,HuardPRXQuantum.2.020323,Eickbusch2021X} and all non-Gaussian operations do require a nonlinear {control} element.

Thus we must confront the central paradox of hybrid bosonic architectures: high-$Q$ resonators offer superior coherence over superconducting qubits, but we cannot achieve universal control of their bosonic modes without utilizing nonlinear elements---such as transmon qubits~\cite{Hofheinz2009,Heeres2017,Kudra2022},  SNAILs~\cite{Eriksson2023X}, or their combination~\cite{Lu2022X}---that have lower coherence. In this paper, we explore the limits on high-fidelity control of bosonic modes with lower fidelity auxiliary nonlinear elements on a simple, but highly relevant problem---exploiting the ultra-high-$Q$ cavity as a memory for a transmon qubit. To understand the magnitude of this challenge, consider that {state-of-the-art} microwave resonators can {currently} be constructed with energy damping times ranging from \SI{35}{\milli\second} \cite{RosenblumCavity2023} for small mode volume cavities to \SI{2}{\second} \cite{Romanenko_PhysRevApplied.13.034032} for large mode volume cavities, while typical transmon superconducting qubits have phase coherence times of only \SIrange{100}{500}{\micro\second} \cite{PlaceTantalum2021,WangTantalum2022} and the best fluxonium qubits~\cite{FluxoniumPhysRevLett.130.267001} have phase coherence times of about \SI{1.5}{\milli\second}. Thus our challenge is to control cavities whose coherence times can be several orders of magnitude larger than those of the required non-linear control elements.

\begin{figure}
    \centering
	\includegraphics[width=\linewidth]{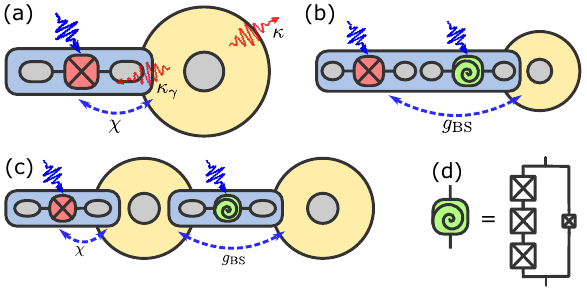}
	\caption{Schematic depiction of memory architectures considered in this paper. (a) A microwave cavity coupled to a transmon qubit via dispersive coupling at a rate $\chi$. Driving both systems (blue wavy arrow for qubit) enables universal cavity control. At the same time, the cavity decays (red arrows) due to its intrinsic losses (rate $\kappa$) and through the qubit due to the inverse Purcell decay (rate $\kappa_\gamma$). (b) To avoid cavity self-Kerr and cross-Kerr interaction with the transmon during idling, the interaction between the cavity and transmon qubit can be mediated by a SNAIL coupler. When the SNAIL coupler is driven at the difference of the cavity and transmon frequencies, a beam-splitter interaction is enabled between them. (c) When the difference in qubit and cavity lifetimes is large, an intermediate buffer cavity can be used. The qubit enacts control on the buffer cavity and the final state is then swapped to the long-lived cavity for storage. (d) Circuit diagram for a SNAIL coupler composed of {three} large serial Josephson junctions connected in parallel to one smaller junction. A dc flux bias can be used to find a suitable operating regime (such as the Kerr-free point where the quartic nonlinearity vanishes, but the cubic non-linearity needed for the drive-induced beam-splitter coupling is still significant).}
 \label{fig:architectures}
\end{figure}

We will organize our approach to this {quantum} engineering challenge around three different hardware architectures that are schematically illustrated in Fig.~\ref{fig:architectures}. Panel (a) illustrates the simplest (and by far the most commonly used) control setup with a transmon qubit directly coupled to a high-$Q$ resonator with cavity control enabled by driving the transmon and/or the cavity field~\cite{Eickbusch2021X,RosenblumCavity2023}; {the most important cavity decoherence channels are highlighted}. Panel (b) illustrates a more complex architecture in which the interaction between the transmon and the resonator is mediated via a three-wave mixing element (a SNAIL~\cite{Frattini2017}) that acts as a microwave-activated switch (beam-splitter interaction) that can rapidly turn the coupling {between the qubit and the cavity} on and off \cite{Ben_Stijn_preprint}. By operating the SNAIL at its Kerr-free point~\cite{SivakSNAIL}, this increased experimental complexity can be used to remove the cross-Kerr interaction between the qubit and cavity {and thereby enhance} cavity coherence {during idling~\cite{Reagor2016}.} Panel (c) adds an intermediate buffer resonator which can provide additional filtering isolation in the case of an extreme mismatch between the qubit and storage cavity coherence lifetimes.

For each of these physical architectures, we will consider different strategies to induce effective interaction terms between the transmon qubit and the storage cavity that can be used to achieve a control rate for executing unitaries on the cavity that is substantially higher than the {decoherence} rate of the transmon.  
Increasing the qubit-cavity (or coupler-cavity) coupling leads to higher control rates but also enhances the cavity decay rate through the inverse Purcell effect \cite{Reagor2016,Zhang2019} and we will explore the design trade-offs that result from this.
Importantly, the inverse Purcell decay rate of the cavity may be significantly larger during the gate operation (where the induced effective-interaction terms are large) than during idle times, and this will need to be taken into account in the design optimization.  The importance of this effect will vary with the memory/compute ratio, but the essence of the engineering challenge is to implement cavity control at a rate faster than the auxiliary qubit coherence decays, while keeping the inverse Purcell decay rate during idling periods low compared to the intrinsic cavity decay rate. If these two conditions are satisfied, the cavity read/write fidelities will be high and the storage time will remain long.

A relevant characteristic for these discussions is the participation ratio \cite{WangParticipationRatio} which quantifies the hybridization between the cavity and the nonlinear circuitry used for its control. Increasing the participation ratio leads to stronger interactions (allowing faster control operations) but also faster cavity decoherence during idling. As we discuss later, this trade-off gives rise to an optimum hybridization for a target idling time loosely akin to the standard quantum limit in cavity optomechanics~\cite{Clerk2010,Aspelmeyer2014}. A key issue is then the scaling of the various control and decoherence rates with the participation ratio which determines whether an ultra-high-$Q$ cavity mode can be efficiently controlled when reducing its participation ratio. This factor represents the main limitation for the case of a cavity directly coupled to a transmon [Fig.~\ref{fig:architectures}(a)] where the inverse Purcell decay is eventually, for small participation ratio, no longer the limiting factor.  Instead the cavity coherence becomes limited by dephasing due to dispersive coupling to thermal noise in the transmon, which depends only on the heating rate of the transmon. This result shows the importance of more sophisticated setups for controlling ultra-high-$Q$ cavity modes, such as the ones shown in Fig.~\ref{fig:architectures}(b,c) or other approaches to protecting the cavity from dispersive interaction during idling~\cite{Noh2023}.

\section{Control and memory fidelity}\label{sec:control}

\subsection{System Hamiltonian and decoherence}

To understand the trade-offs associated with controlling ultra-high-$Q$ microwave cavities using lower-$Q$ nonlinear circuits, let us assume a cavity mode $a$ coupled to a qubit mode $q$ at a rate $g$ via the interaction Hamiltonian
\begin{equation}\label{eq:Hint}
	H_{\rm int} = g (a^\dagger q + q^\dagger a),
\end{equation}
which, along with microwave drives on the cavity and qubit, allows universal control of the cavity mode using a range of different techniques~\cite{Hofheinz2009,Heeres2017,Eickbusch2021X,Kudra2022}. Some of these strategies rely on {a tunable} resonant interaction (where the frequencies of the cavity and qubit are equal, $\omega_a = \omega_q$), while others exploit dispersive coupling (available in the large-detuning regime, $\Delta = \omega_a-\omega_q\gg g$) of the form
\begin{equation}\label{eq:Hdisp}
    H_{\rm disp} = -\chi a^\dagger aq^\dagger q.
\end{equation}
This dispersive interaction relies on the fourth-order nonlinearity of the qubit, $H_4 = -\frac{1}{2}Kq^{\dagger 2}q^2$, for which $\chi\approx 2K(g/\Delta)^2$ and which introduces an additional cavity self-Kerr term $H_{\rm sK} = -K_aa^{\dagger 2}a^2$ with $K_a\approx \frac{1}{2}K(g/\Delta)^4$; the most commonly used such qubit is the transmon. Alternatively, we will consider third-order nonlinearity as is typical in SNAIL devices, $H_3 = g_3(q+q^\dagger)^3$ \cite{Frattini2017}. With these circuits, instead of using the dispersive interaction given in Eq.~\eqref{eq:Hdisp}, cavity control is enabled by three-wave mixing that relies on the cubic nonlinearity of the SNAIL (we will use this to engineer Gaussian transformations; see Appendix~\ref{sec:mixing} for a general description of three- and four-wave mixing processes in nonlinear circuits and analysis of their performance). One might now expect that efficient control is a question of sufficiently strong coupling but this assumption omits the decoherence channels inherited by the (high-$Q$) cavity from the (comparatively low-$Q$) qubit.

The hybridization of the cavity with the qubit introduces an additional decay channel for the cavity mode. The precise description of this decoherence depends on the specific experimental arrangement and error channels of the qubit; most commonly, the interaction introduces an additional cavity loss (or thermal noise in the presence of thermal qubit excitations), dephasing, and dressed dephasing {modeled by} Lindblad operators combining cavity and qubit operators~\cite{BlaiscQEDReviewRMP2020}; see also Appendix~\ref{sec:masterEq}. The additional cavity loss is given by the inverse Purcell decay rate which varies between
\begin{equation}
    \kappa_\gamma = \frac{\kappa+\gamma}{2}
\end{equation}
for {the} resonant Jaynes--Cummings interaction (here, $\kappa$ and $\gamma$ are the relaxation rates of the cavity and qubit, respectively) and
\begin{equation}
    \kappa_\gamma = \gamma\left(\frac{g}{\Delta}\right)^2
\end{equation}
for the dispersive regime. Although an approximate formula interpolating between these limits can be derived~\cite{Barends2013}, these are the only two regimes relevant for the rest of our paper. Clearly, increasing the dispersive coupling $\chi$ by enhancing the cavity participation ratio $(g/\Delta)^2$ also increases the inverse Purcell decay $\kappa_\gamma$, eventually destroying the advantage of long intrinsic cavity lifetime. In the following, we will therefore limit the inverse Purcell decay (by constraining the participation ratio) to $\kappa_\gamma\lesssim\kappa$. We will call the situation $\kappa_\gamma = \kappa$ critical inverse Purcell decay or critical coupling to the qubit in analogy with critical coupling of localized modes to their inputs/outputs.

Additional cavity dephasing is caused by thermal jumps of the qubit at rates $\gamma_\uparrow = \gamma\bar{n}_q$ (ground to excited), $\gamma_\downarrow = \gamma(\bar{n}_q+1)$ (excited to ground), where $\bar{n}_q\ll 1$ is the thermal population of the qubit. In the dispersive regime, these jumps randomly shift the cavity resonance frequency via the cross-Kerr interaction, giving rise to effective dephasing; in the strong coupling regime, $\chi\gg\gamma$, its rate can be approximated as $\gamma\bar{n}_q$~\cite{Reagor2016}. The qubit dephasing, caused by random frequency fluctuations of its transition, modifies the cavity Lamb shift, resulting in cavity dephasing that is quadratic in the participation ratio $(g/\Delta)^2$,
\begin{equation}
    \kappa_{\gamma_\phi} = \gamma_\phi\left(\frac{g}{\Delta}\right)^4,
\end{equation}
although a deviation from this scaling might occur due to low-frequency pink noise~\cite{Ben_Stijn_preprint}.
Finally, the dressed modes in the dispersive regime lead to the existence of dressed dephasing at a rate
\begin{equation}
    \gamma_\Delta = \gamma_{\phi E}\left(\frac{g}{\Delta}\right)^2
\end{equation}
associated with Lindblad operators $a^\dagger q,q^\dagger a$~\cite{BlaiscQEDReviewRMP2020}, where $1/\gamma_{\phi E}$ is the echoed qubit dephasing time (see Table~\ref{tab:parameters}). We take the echoed dephasing rate as an upper bound since the relevant noise is at frequency $\Delta$ and therefore the $1/f$ noise in the pure dephasing rate $\gamma_\phi$ would give a too pessimistic estimate; we will also assume that the noise density at frequency $\Delta$ and $-\Delta$ is the same so we can use the same rate $\gamma_\Delta$ for both dressed dephasing terms.

The total master equation describing the dressed dynamics in the dispersive regime is (see Appendix~\ref{sec:masterEq} for full derivation and explanation of certain small terms that have been neglected)
\begin{equation}\label{eq:master_main}
\begin{split}
    \dot{\rho} &= -i[H_{\rm int}+H_{\rm dr},\rho] + \mathcal{L}_{{\rm th},a}(\kappa,\bar{n}_a)\rho + \mathcal{L}_{{\rm th},a}(\kappa_\gamma,\bar{n}_q)\rho \\
    &\quad + \mathcal{L}_{{\rm th},q}(\gamma,\bar{n}_q)\rho + (\kappa_\phi + \kappa_{\gamma_\phi})\mathcal{D}[a^\dagger a]\rho + \gamma_\phi\mathcal{D}[q^\dagger q]\rho \\
    &\quad + \gamma_\Delta\mathcal{D}[a^\dagger q]\rho + \gamma_\Delta\mathcal{D}[q^\dagger a]\rho \\
    &\quad + \sqrt{\gamma_\phi\kappa_{\gamma_\phi}}(\mathcal{S}[a^\dagger a,q^\dagger q] + \mathcal{S}[q^\dagger q, a^\dagger a])\rho.
\end{split}
\end{equation}
In this equation, we use the total interaction Hamiltonian
\begin{equation}
    H_{\rm int} = -\frac{1}{2}Kq^{\dagger 2}q^2-\chi a^\dagger aq^\dagger q - K_a a^{\dagger 2}a^2 - \chi' a^{\dagger 2}a^2q^\dagger q
\end{equation}
with $\chi' = 2K(g/\Delta)^4$ representing the non-linearity of the dispersive shift, and the general Hamiltonian $H_{\rm dr}$ describing the driving of the cavity and qubit. While corrections to the dispersive shift from the $\chi'$ term are often negligible at low photon numbers, when populating the oscillator with many photons this term must be accounted for in order to reach high-fidelity control \cite{Eickbusch2021X}. Thermal decoherence associated with different operators $o=q,a$ in Eq.~(\ref{eq:master_main}) is described by the super-operator
\begin{equation}
    \mathcal{L}_{{\rm th},o}(\Gamma,\bar{n})\rho = \Gamma(\bar{n}+1)\mathcal{D}[o]\rho + \Gamma\bar{n}\mathcal{D}[o^\dagger]\rho
\end{equation}
and we assume different thermal occupations for the cavity ($\bar{n}_a$) and transmon ($\bar{n}_{q}$) baths (note that the cavity couples to the transmon bath via the inverse Purcell effect). In a 3D superconducting cavity architecture, this is a reasonable assumption, as the cavity is often at lower temperature than the qubit due to better thermal anchoring and a lower susceptibility to anomalous excitation from quasiparticle poisoning. We use the standard Lindblad term $\mathcal{D}[o]\rho = o\rho o^\dagger -\frac{1}{2}(o^\dagger o\rho+\rho o^\dagger o)$ and the non-Lindblad super-operators $\mathcal{S}[o_1,o_2]\rho = o_1\rho o_2^\dagger - \frac{1}{2}(o_2^\dagger o_1\rho + \rho o_2^\dagger o_1)$~\cite{Eickbusch2021X}. 

One limitation of this approach to qubit-induced cavity decoherence is that, while it properly accounts for absorption of cavity photons by the damped qubit resonance transition, it does not account for other non-resonant effects associated with the qubit chip such as microwave absorption by defects on the surface and in the bulk of the chip. These mechanisms are not fully understood at present so we do not include them in our numerical simulations; however, we provide a simple quantitative estimate on their size relative to the inverse Purcell decay in Sec.~\ref{sec:losses}.

\subsection{Trade-off between control and storage}

For a good quantum memory, efficient control and storage of quantum states require a compromise between fast gates and slow decay. Depending on the intended storage time (which may or may not be known in advance), fast control (enabled by strong coupling to the qubit) can be detrimental as it leads to stronger inverse Purcell decay (and other dressed dissipation processes). On the other hand, for short storage times, a strongly under-coupled cavity ($\kappa_\gamma\ll\kappa$) unnecessarily limits the available coupling and gate fidelity. The overall performance of the memory {should thus} be characterized by the memory fidelity, which combines the fidelity of control and errors accumulated during storage~\cite{Girvin_LesHouches_QEC}.

Here, using the dispersive architecture depicted in Fig.~\ref{fig:architectures}(a), we will consider the case of storing a single qubit for time $t_\text{write} + t_i + t_\text{read}$ by writing the information from the auxiliary qubit into the vacuum and single-photon states of an ultra-high-$Q$ oscillator (taking time $t_{\rm write}$), waiting for a time $t_i$, and reading the information out, back to the auxiliary qubit (time $t_{\rm read}$). For a particular state $\ket{\psi}$, the fidelity of this process is given by the memory fidelity $\mathcal{F}_m = \langle\psi|\rho_{\rm actual}|\psi\rangle$, where $\rho_{\rm actual}$ is the actual (mixed) state we obtain as a result of applied control and decoherence. To evaluate the capability of the memory, we further average over a set of intended states (i.e., over the whole qubit Bloch sphere), which gives us the average memory fidelity $\bar{\mathcal{F}}_m$~\cite{Nielsen2000}. For a detailed analysis of the whole process, we will approximate the average memory fidelity by $\bar{\mathcal{F}}_m \approx \bar{\mathcal{F}}_c \bar{\mathcal{F}}_i$, where $\bar{\mathcal{F}}_c$ is the \emph{control} fidelity, accounting for error during the read and write operations, while $\bar{\mathcal{F}}_i$ is the \textit{idling} fidelity, accounting for error during storage in the cavity. In this approach, we include only incoherent errors (stemming from decoherence) and do not consider coherent unitary errors (such as over- or under-rotations due to microwave pulse distortion).

For dispersive control, the gate time during control (for write in and read out of the cavity) is inversely proportional to the dispersive coupling, {$ t_\text{read} + t_\text{write} \approx  \alpha /\chi$, where the dimensionless parameter $\alpha$ accounts for the fact that we might need several gates of different length to effect the target operation. As a result,} the control fidelity $\bar{\mathcal{F}}_c$ is limited primarily by qubit decoherence, $\bar{\mathcal{F}}_c = 1-\epsilon_c \sim 1-\alpha\gamma_{\rm tot}/\chi$, where $\gamma_{\rm tot}$ is the total dissipation rate of the qubit, involving transmon relaxation, heating, and dephasing. (We do not include a precise formula here since relaxation and dephasing affect the dynamics differently and the specifics depend on the particular control sequence. We will consider several possible such situations in Sec.~\ref{sec:memory_sim} and derive the corresponding formulas then.) The other dissipation processes during control can be neglected since we assume a much higher $Q$ for the cavity, $\kappa\ll\gamma$, $\kappa_\phi\ll\gamma_\phi$, and validity of the dispersive approximation, $(g/\Delta)^2\ll 1$ which limits dressed dissipation [i.e., dissipation processes with rates involving powers of $(g/\Delta)^2$]. Relaxation, heating, and dephasing generally affect quantum operations differently but combining them in a single decoherence rate $\gamma_{\rm tot}$ allows us to approximate their combined effect using a simple formula.

During idling in the cavity, errors are caused by the intrinsic cavity dissipation and its dressing by the qubit, $\bar{\mathcal{F}}_i = 1-\bar{\epsilon}_i = 1-\bar{\epsilon}_0-\bar{\epsilon}_d$. The intrinsic dissipation gives the total error $\bar{\epsilon}_0 \sim \kappa_{\rm tot}nt_i$ for idling time $t_i$ with total cavity dissipation rate $\kappa_{\rm tot}$ (including again relaxation, heating, and dephasing) and average photon number $n$ (which varies for the different qubit states being averaged over), while the leading-order error due to dressed dissipation is $\bar{\epsilon}_d \sim \bar{\gamma}_{\rm tot}n(g/\Delta)^2 t_i$ (where $\bar{\gamma}_{\rm tot}$ can in general differ from $\gamma_{\rm tot}$ used to characterize control errors). The dressed error stems from the inverse Purcell decay and heating of the cavity and the dressed dephasing with jump operator $q^\dagger a$.
The conjugate dressed dephasing term $a^\dagger q$ contributes only in higher order due to thermal excitation of the qubit as do the non-Lindblad terms in the last line of Eq.~\eqref{eq:master_main}. The qubit-induced dephasing of the cavity at rate $\gamma_\phi(g/\Delta)^4$ can be neglected since it is of higher power in the participation ratio $(g/\Delta)^2$.

Approximating the participation ratio in terms of the qubit anharmonicity and dispersive coupling, $(g/\Delta)^2 \approx \chi/2K$, we obtain the total average memory fidelity for the dispersive case
\begin{equation}
    \bar{\mathcal{F}}_m = \bar{\mathcal{F}}_c\bar{\mathcal{F}}_i \sim 1 - \frac{\alpha\gamma_{\rm tot}}{\chi} - \kappa_{\rm tot}nt_i - \frac{\bar{\gamma}_{\rm tot}\chi}{2K}nt_i,
\end{equation}
where we assume small errors for all three fidelities, $1-\bar{\mathcal{F}}_j\ll 1$, $j=c,i,m$. If the idling time is known in advance (for example if one is designing short-term or long-term memory), we can find the optimal coupling by making the $\chi$-dependent errors equal, $\bar{\epsilon}_c = \bar{\epsilon}_d$, which leads to
\begin{equation}
    \chi_{\rm opt} = \sqrt{\frac{2\alpha\gamma_{\rm tot} K}{\bar{\gamma}_{\rm tot} nt_i}} \simeq \sqrt{\frac{2\alpha K}{nt_i}},
\end{equation}
where we assumed $\gamma_{\rm tot}\simeq \bar{\gamma}_{\rm tot}$ for simplicity. The memory fidelity then simplifies to
\begin{equation}
    \bar{\mathcal{F}}_m^{\rm opt} \sim 1 - \gamma_{\rm tot}\sqrt{\frac{2\alpha nt_i}{K}} - \kappa_{\rm tot}nt_i.
\end{equation}

Throughout this analysis, we have made a few key assumptions. First, limiting the discussion to the vacuum and single-photon states allowed us to neglect the effects of cavity self-Kerr interaction, which would otherwise introduce additional dephasing of the cavity state. Moreover, considering multi-photon cavity states would lead to faster damping and heating of the cavity (due to both intrinsic and inverse Purcell dissipation) enhanced by the average photon number $n$. Note also that for some control techniques based on the dispersive interaction (such as sideband transitions~\cite{Murch2012,RosenblumCavity2023} or conditional displacements~\cite{Eickbusch2021X,diringer2023X}), the control rate exhibits (thanks to relying on four-wave mixing or enhancing the interaction by a large intracavity amplitude) a different scaling with the participation ratio $(g/\Delta)^2$, giving modified expressions for control fidelity and optimal coupling.

Finally, we have not considered the possibility of error correction on the encoded state while idling. Since our discussion is quite general, we expect the scaling to hold also for measuring error syndromes and applying correction operations. These steps, as seen by the cavity field, represent an additional gate sequence performed using one of the available control techniques with a little extra time needed to measure the error syndrome encoded in the qubit state and process the classical result to implement a suitable correction. A key task is then to optimize the idling time between two rounds of error correction. We can expect a similar analysis to hold for this optimization, if the overhead of utilizing the higher-excited cavity states is accounted for.

\section{Uncorrected cavity memory}

\subsection{Simulations of memory fidelity}\label{sec:memory_sim}

\begin{table}
    \centering
    \caption{Parameters used in numerical simulations. The thermal cavity population corresponds to \SI{3.8}{\giga\hertz} at a temperature of \SI{20}{\milli\kelvin}. Formulas involving the detuning 
    $\Delta$ assume it is large relative to the charging energy of the transmon, $\Delta\gg E_c\simeq K$.  The dimensionless pump strength $\xi_1$ is defined in Appendix~\ref{sec:mixing}. Consistent with recent experiments~\cite{Ben_Stijn_preprint,GaoYY2018}, we use stronger driving for three-wave mixing than four-wave mixing since it produces fewer spurious induced resonances, although new techniques exist 
    that solve this problem and can create strong four-wave mixing via flux modulation~\cite{YaoLu_preprint}.}\label{tab:parameters}
\begin{tabular}{|l|cc|}
	\hline
    Parameter & Symbol & Value/expression \\
    \hline \hline
	Transmon Kerr nonlinearity & $K/(2\pi)$ & \SI{200}{\mega\hertz} \\
	SNAIL cubic nonlinearity & $g_3/(2\pi)$ & \SI{5}{\mega\hertz} \\
	Qubit relaxation time & $1/\gamma$ & \SI{200}{\micro\second} \\
    Qubit thermal population & $\bar{n}_q$ & $10^{-3}$ \\
	Qubit pure dephasing time & $1/\gamma_\phi$ & \SI{400}{\micro\second} \\
    Qubit echoed dephasing time & $1/\gamma_{\phi{}E}$ & \SI{1000}{\micro\second} \\ 
	Cavity thermal population & $\bar{n}_{\rm cav}$ & $10^{-4}$ \\
    Cavity pure dephasing time & $1/\kappa_\phi$ & \SI{500}{\milli\second} \\
    Pump strength (three-wave) & $\xi_1$ & 2 \\
    Pump strength (four-wave) & $\xi_1$ & 0.1 \\
	\hline \hline
    Dispersive coupling rate & $\chi$ & $2K(g/\Delta)^2$ \\
    Second-order dispersive shift & $\chi'$ & $2K(g/\Delta)^4$ \\
    Beam-splitter coupling rate & $g_{\rm BS}$ & $6g_3\xi_1(g_a/\Delta_a)(g_b/\Delta_b)$ \\
    Sideband transition rate & $g_{\rm sb}$ & $\frac{1}{2}K\xi_1(g/\Delta)$ \\
    Cavity self-Kerr strength & $K_a$ & $\frac{1}{2}K(g/\Delta)^4$ \\
    Inverse Purcell decay rate & $\kappa_\gamma$ & $\gamma(g/\Delta)^2$ \\
    Inverse Purcell dephasing rate & $\kappa_{\gamma_\phi}$ & $\gamma(g/\Delta)^4$ \\
    Dressed dephasing rate & $\gamma_\Delta$ & $\gamma_{\phi E}(g/\Delta)^2$ \\
    \hline \hline
    Cavity relaxation time & $1/\kappa$ & \SI{10}{\milli\second}--\SI{10}{\second} \\
    \hline
\end{tabular}
\end{table}

To expand on the previous analysis and obtain meaningful quantitative results on attainable memory fidelities, we now consider alternative architectures (shown in Fig.~\ref{fig:architectures} and discussed above) and control approaches for a quantum memory using the vacuum and single-photon states of an ultra-high-$Q$ cavity. Using numerical simulations, we compare these schemes to the simple case of keeping the qubit state in the ground and first excited state of the transmon, $\ket{\psi} = \alpha\ket{g}+\beta\ket{e}$. We want to quantify the improvement in average memory fidelity (gained by writing the qubit state into the cavity, storing it there for time $t_i$, and reading it out back into the transmon) and estimate the potential benefit of such memories for practical quantum algorithms. For numerical simulations, we consider parameters corresponding to state-of-the-art superconducting devices (with the exception of cavity lifetimes which are assumed up to \SI{10}{\second} long) as listed in Table~\ref{tab:parameters}.

\begin{figure}
    \centering
	\includegraphics[width=1.0\linewidth]{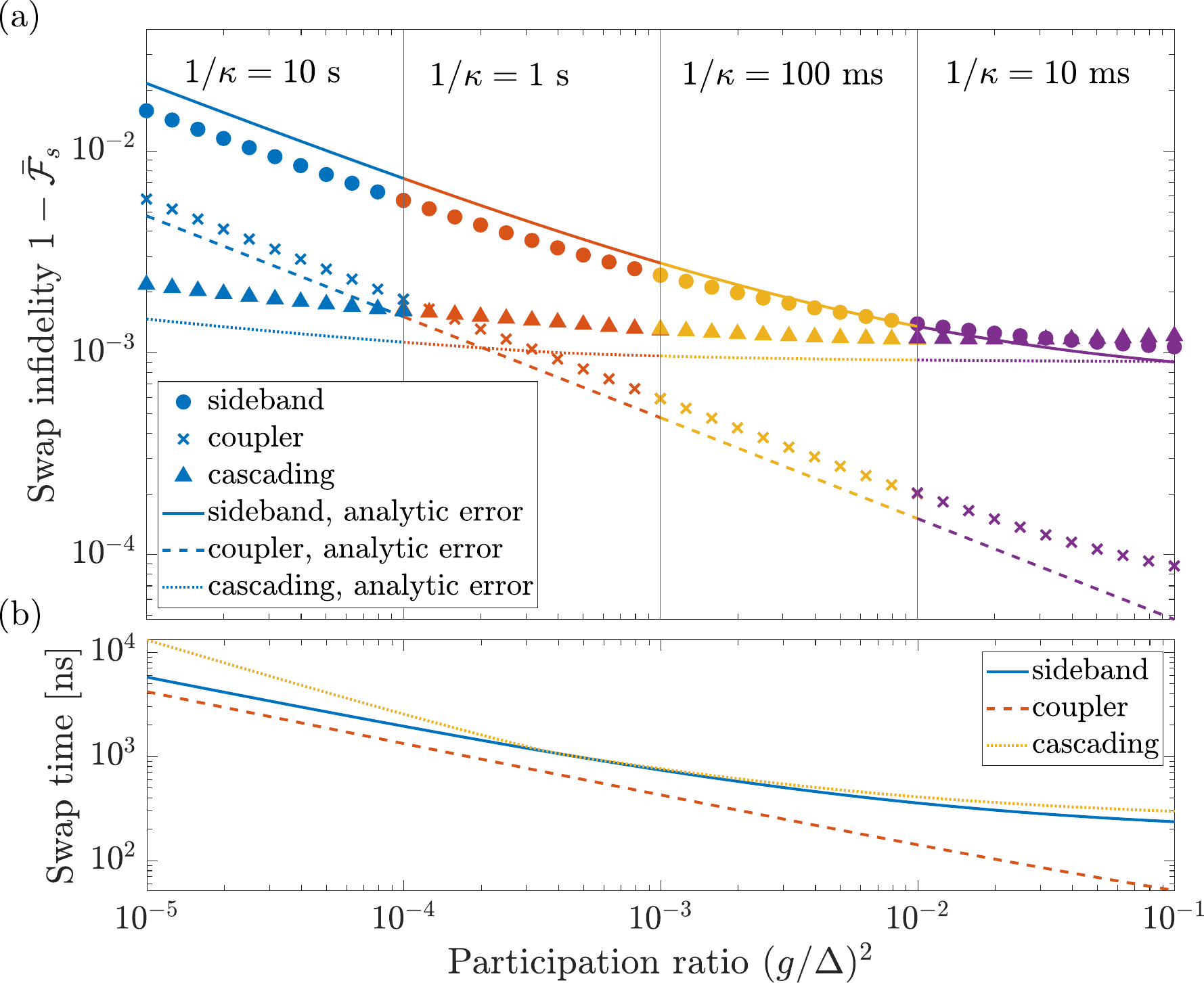}
	\caption{(a) Comparison of the channel infidelities for swapping a qubit state from the $g$--$e$ manifold of a transmon into an ultra-high-$Q$ cavity (encoded in the vacuum and single photon). The simulations are done with the direct coupling sideband drive (transmon-cavity, circles), SNAIL coupler with resonant beam-splitter interaction (transmon-coupler-cavity, crosses), and cascaded cavities (transmon-buffer-coupler-cavity, triangles). The lines are analytically calculated errors using Eqs.~\eqref{eq:error_SDB}, \eqref{eq:error_BS} (solid line for sideband drive, dashed for SNAIL coupler, dotted for cascaded cavities). The simulations are done with four different cavity lifetimes ranging from $1/\kappa = \SI{10}{\second}$ down to $1/\kappa = \SI{10}{\milli\second}$ as indicated at the top of the plot with different lifetime in each of the four panels and different color lines and data points. (The intrinsic lifetime for data points at the boundaries between the panels is determined only by the color; for swap, this makes no difference since errors are dominated by the transmon decoherence.) For each value of $\kappa$, the inverse Purcell rate $\kappa_\gamma$ ranges from $\frac{1}{2}\kappa$ to $5\kappa$. For the buffer cavity in cascading, we take the intrinsic decay rate to be the geometric mean of the transmon and storage cavity, $\kappa_b = \sqrt{\kappa\gamma}$. The other parameter values are displayed in Table~\ref{tab:parameters}. (b) The duration of the swap operation for the numerical data shown in (a). The swap time includes smooth ramping up and down of the driving tones.}\label{Fig:SWAP}
\end{figure}

As the basic building block for evaluating the memory fidelity, we consider the swap fidelity $\bar{\mathcal{F}}_s$, i.e., the fidelity of the write operation. Due to the (ideal) write and read operations being Hermitian conjugates of each other, we can reasonably assume that the read fidelity will be equal to the write fidelity and the total control fidelity can be approximated as $\bar{\mathcal{F}}_c \approx \bar{\mathcal{F}}_s^2$. For each of the three architectures in Fig.~\ref{fig:architectures}, we perform a suitable swap operation as described in general in the following with technical details deferred to Appendix~\ref{sec:MEqs}. In the case of the transmon directly coupled to the cavity [Fig.~\ref{fig:architectures}(a)], we use sideband driving which allows direct transitions via the transmon's  four-wave mixing~\cite{RosenblumCavity2023} between the states $\ket{f,0}$ and $\ket{g,1}$, where $\ket{f}$ is the second excited state of the transmon. Since we assume that the qubit state is initially encoded in the $g$--$e$ manifold, the sideband transition has to be preceded by a $\pi$ pulse on the $e$--$f$ manifold. As can be seen in Fig.~\ref{Fig:SWAP}(a) (circles), the $e$--$f$ rotation represents a constant error of about 0.1 \%. This is the main limiting factor for systems with large participation ratio $(g/\Delta)^2$ where the subsequent sideband transition at a rate $g_{\rm sb} = \frac{1}{2}K\xi_1(g/\Delta)$ is fast. For small participation ratios, the swap fidelity is limited by the transmon decoherence during the sideband transition, giving rise to the $(g/\Delta)^{-1}$ scaling of the total swap error.

These features can be described by a simple analytical model assuming at most one quantum jump during the whole swap operation ($e$--$f$ rotation and sideband transition)~\cite{Abad2022}. As we describe in detail in Appendix~\ref{sec:fidelities}, the errors can be approximated as
\begin{subequations}\label{eq:error_SDB}
\begin{align}
    \bar{\epsilon}_{ef} &= {1-\bar{\mathcal{F}}_{ef} \approx} \frac{7}{12}\gamma t_{ef} + \frac{11}{24}\gamma_\phi t_{ef},\\
    \bar{\epsilon}_{\rm sb} &= {1-\bar{\mathcal{F}}_{\rm sb} \approx} \frac{\gamma+\gamma_\phi}{2}t_{\rm sb},
\end{align}
\end{subequations}
where $t_{ef}$ is the gate time for the $e$--$f$ rotation (set by the amplitude of the qubit drive and independent of $g/\Delta$), and $t_{\rm sb} = \pi/(2g_{\rm sb})$ is the time for the sideband transition, which is inversely proportional to $g/\Delta$. The total swap fidelity is then given by the product of the fidelities of the $e$--$f$ rotation and sideband transition, $\bar{\mathcal{F}}_s \simeq \bar{\mathcal{F}}_{ef}\bar{\mathcal{F}}_{\rm sb}\simeq 1-\bar{\epsilon}_{ef}-\bar{\epsilon}_{\rm sb}$. This simple model, which assumes that errors in both steps are independent and caused primarily by relaxation and dephasing of the transmon (neglecting transmon heating and all cavity decoherence processes), gives good agreement with the numerical simulations.

Alternatively, the transmon and cavity can be coupled via a SNAIL coupler as shown in Fig.~\ref{fig:architectures}(b). Pumping the SNAIL at the difference of the transmon and cavity frequencies results in their beam-splitter interaction (see Appendix~\ref{sec:mixing} for details), which again allows the transmon state to be swapped to the cavity. This approach allows swapping the qubit state directly from the $g$--$e$ manifold without the need to first transfer the population from the first to the second excited state. Since the beam-splitter rate $g_{\rm BS} = 6g_3\xi_1(g_q/\Delta_q)(g/\Delta)$ (see Appendix~\ref{sec:mixing}) is also linearly dependent on the cavity participation ratio (now with the coupler and not the transmon), we get the same scaling of swap error as for the sideband driving in the small $(g/\Delta)^2$ limit [crosses in Fig.~\ref{Fig:SWAP}(a)]. [The other parameters in $g_{\rm BS}$ are the SNAIL cubic nonlinearity $g_3$, normalized drive strength $\xi_1$, and transmon-coupler participation ratio $(g_q/\Delta_q)^2$; to keep the coupling rate as large as possible while maintaining all approximations, we use $(g_q/\Delta_q)^2 = 0.1$.] The beam-splitter error is then (see Appendix~\ref{sec:fidelities} for derivation)
\begin{equation}\label{eq:error_BS}
    \bar{\epsilon}_{\rm BS} = {1-\bar{\mathcal{F}}_{\rm BS}=} \frac{1}{6}\gamma t_{\rm BS} + \frac{1}{8}\gamma_\phi t_{\rm BS};
\end{equation}
since the gate time is, for our parameters, shorter than for the sideband transition, $t_{\rm BS} = \pi/(2g_{\rm BS})<t_{\rm sb}$ [see Fig.~\ref{Fig:SWAP}(b)], and the prefactors are smaller, the coupler allows a smaller error than the sideband transition. In addition, as the SNAIL can be operated at the Kerr-free point, it can be used to suppress thermally induced dephasing of the cavity during idling as we will discuss later.

Finally, we consider write/read operations performed with the help of a buffer cavity as shown in Fig.~\ref{fig:architectures}(c). This buffer cavity---while potentially having the same intrinsic $Q$ factor as the storage cavity---can have much larger participation in the transmon and SNAIL coupler, thereby allowing fast extraction of the state from the transmon while maintaining good Purcell protection of the ultra-high-$Q$ storage cavity during idling. We assume that the sideband driving scheme is used to swap the qubit state from the transmon to the buffer cavity, followed by a beam-splitter interaction between the buffer and storage cavities. With the help of the analytical estimates of errors in the various steps, Eqs.~\eqref{eq:error_SDB}, \eqref{eq:error_BS}, we can optimize the participation of the buffer cavity in both transmon and coupler for each value of the participation ratio of the storage cavity.

\begin{figure}
\begin{center}
	\includegraphics[width=1.0\linewidth]{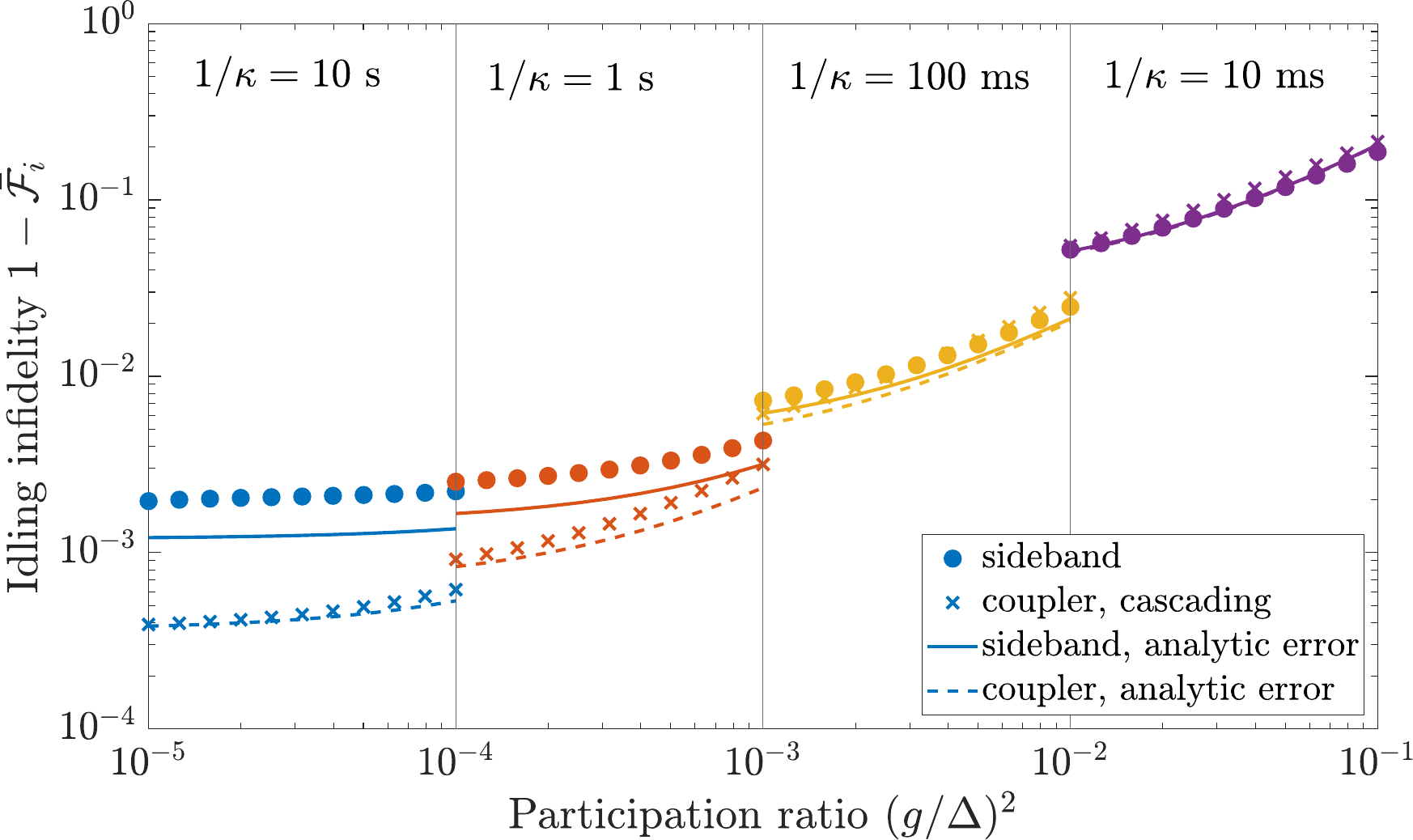}
	\caption{Comparison of the channel infidelities of idling ($t_i = \SI{1}{\milli\second}$) a Fock-encoded qubit state in an ultra-high-$Q$ cavity. Coupler and cascaded cavities have the same idling fidelity so only the former is plotted here. The different curves and simulation parameters are the same as in Fig.~\ref{Fig:SWAP}. }\label{Fig:IDLE}
\end{center}
\end{figure}

The results, plotted using triangles in Fig.~\ref{Fig:SWAP}(a), show the combined effect of these errors: For large participation ratio (of the storage cavity), the performance is limited by the $e$--$f$ rotation of the transmon (followed by high-fidelity sideband transition and beam-spliter operations). As the participation ratio is reduced, the error increases slower than for the other two techniques thanks to the buffer cavity and its additional optimization parameters (participation ratios of the buffer cavity in the transmon and coupler). For very small participation ratios [leftmost panel in Fig.~\ref{Fig:SWAP}(a)], cavity cascading outperforms even the coupler scheme due to a relatively fast swap of the qubit state from the transmon to the buffer cavity; since the buffer has a much longer lifetime than the transmon, the subsequent swap from the buffer to the storage cavity can be much slower while maintaining high fidelity. In these simulations, we take the intrinsic damping rate of the buffer cavity as the geometric mean of the transmon and storage cavity decay rates. The buffer cavity does not need the same ultra-high $Q$ as the storage cavity due to its much faster inverse Purcell decay (into both transmon and SNAIL); we choose the value $\kappa_b = \sqrt{\kappa\gamma}$ as a suitable lower bound on the intrinsic lifetime that preserves all the advantages discussed here.

Having understood the main limitations on the swap fidelity, we now turn our attention to idling, which is analyzed in Fig.~\ref{Fig:IDLE}. The coupler and cavity cascading perform equally due to identical error channels (the same intrinsic relaxation and dephasing and additional decoherence due to SNAIL), but the cavity directly coupled to a transmon shows larger error for small $(g/\Delta)^2$. To understand this difference, we can evaluate the idling error for time $t_i$~\cite{Abad2022},
\begin{equation}
    \bar{\epsilon}_i = {1-\bar{\mathcal{F}}_i =} \frac{1}{6}(2\Gamma_\downarrow + \Gamma_\phi) t_i,
\end{equation}
where $\Gamma_{\downarrow,\phi}$ denote the total relaxation and dephasing rates for the cavity. For all three setups, the total relaxation is given by the intrinsic and inverse Purcell decay, $\Gamma_\downarrow = \kappa+\kappa_\gamma$. For strong hybridization between the cavity and transmon/coupler [large $(g/\Delta)^2$], the fidelity is limited by this relaxation, $\Gamma_\downarrow\gg\Gamma_\phi$, and all setups have similar performance. Reducing the hybridization decreases $\Gamma_\downarrow$ and leads to a stronger effect of dephasing (relative to relaxation), resulting in worse performance of the direct coupling approach. The total dephasing rate includes the intrinsic and inverse Purcell dephasing processes (at rates $\kappa_\phi$ and $\kappa_{\gamma_\phi}$) for all three schemes. However, for directly coupled transmon and cavity, additional dephasing is caused by thermal noise in the transmon which, due to their cross-Kerr interaction, causes a frequency-shift for the cavity whenever the transmon emits or absorbs a photon from its bath. Since this process is faster than the other dephasing processes (it occurs at a rate $\gamma\bar{n}_q$~\cite{Reagor2016}), it leads to a larger idling error, although note that it can be counteracted using transmon measurement and feedback~\cite{Goldblatt2024}.

\begin{figure}
\begin{center}
	\includegraphics[width=1.0\linewidth]{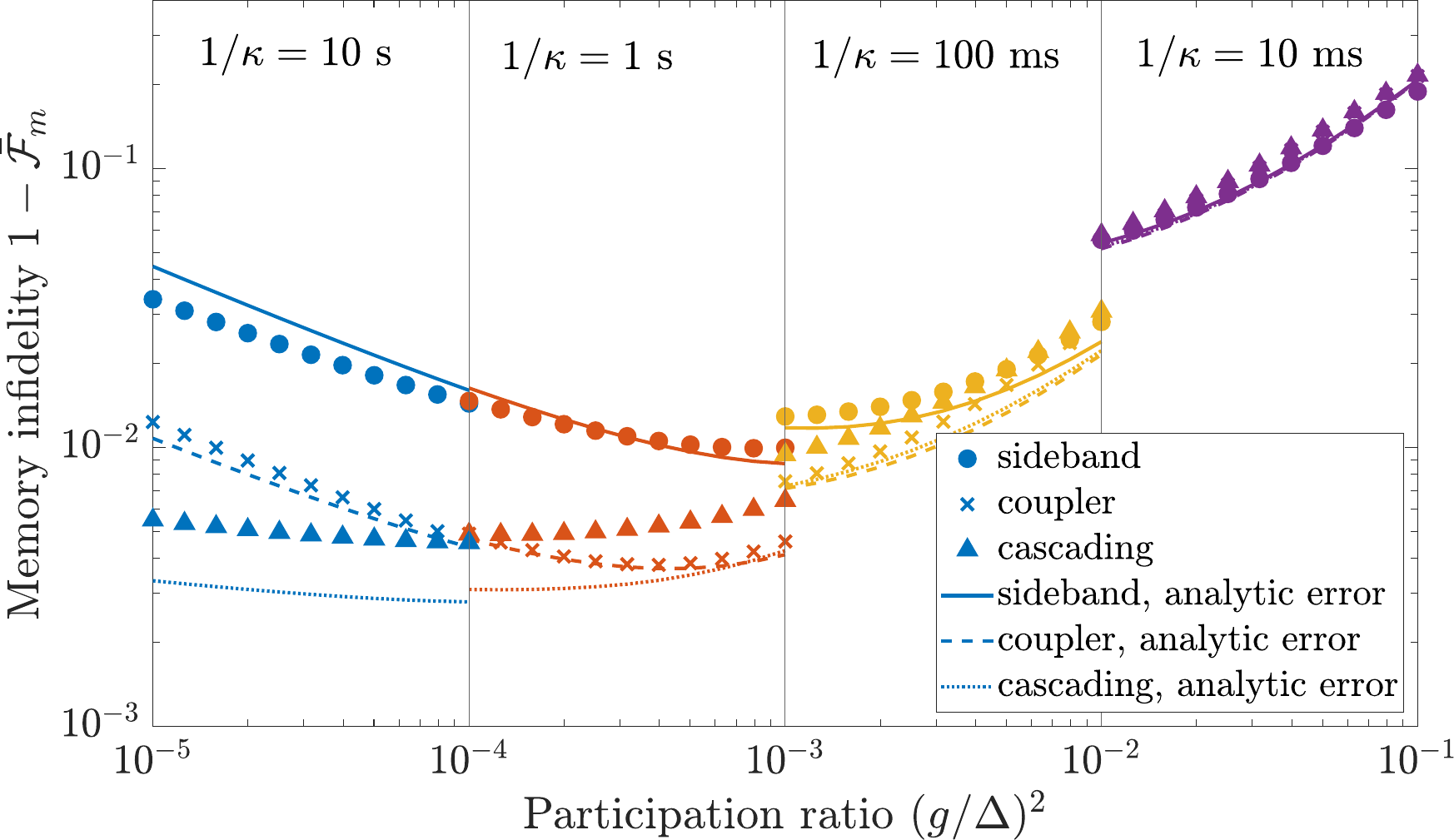}
	\caption{Comparison of the total memory channel infidelities, consisting of swapping an initial qubit state from the transmon (initially in the $g$--$e$ manifold) to the storage cavity, idling for 1 millisecond, and swapping the state back to the transmon using the different architectures shown in Fig.~\ref{fig:architectures}. The different curves and simulation parameters are the same as in Fig.~\ref{Fig:SWAP}.}\label{Fig:SWAPIDLE}
\end{center}
\end{figure}

\begin{figure}
\begin{center}
	\includegraphics[width=1.0\linewidth]{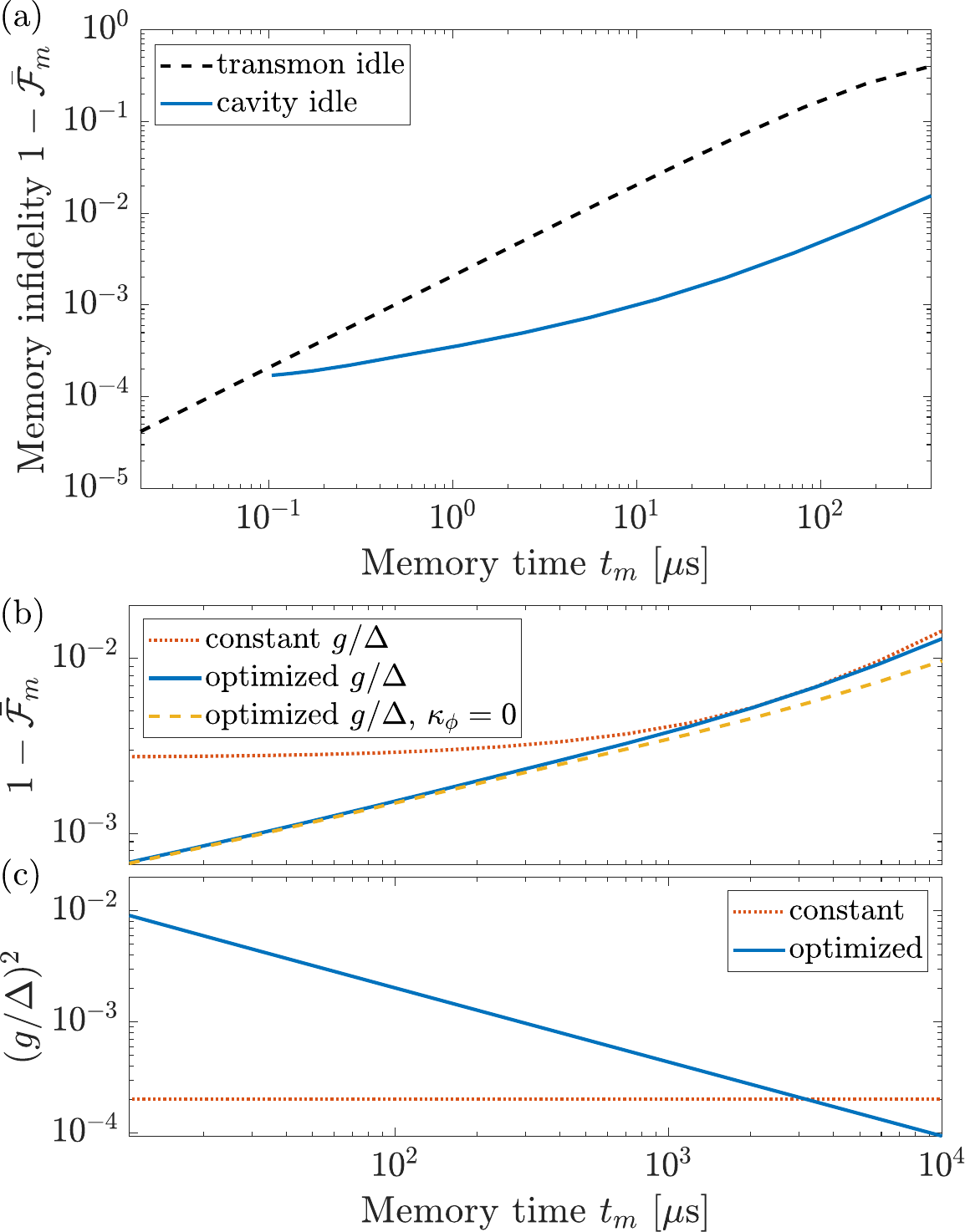}
	\caption{Memory infidelity as a function of memory time $t_m$ for the SNAIL coupler setup. (a) Comparison of cavity memory with $1/\kappa = \SI{10}{\milli\second}$ lifetime (solid blue line) and idling in the transmon (dashed black). For the cavity memory, the participation ratio $(g/\Delta)^2$ is optimized for each value of $t_m$. The shortest memory time is set by the speed of the beam-splitter operation ({$2t_{\rm BS} = \SI{103}{\nano\second}$}) which corresponds to the leftmost point of the curve. (b) Memory infidelity for longer storage times (for cavity memory with $1/\kappa = \SI{1}{\second}$) with constant (dotted red line, corresponding to critical inverse Purcell decay, $\kappa = \kappa_\gamma$) and optimized (solid blue) participation ratio. We also show memory infidelity for numerically optimized participation ratio without intrinsic cavity dephasing, $\kappa_\phi = 0$ (dashed orange); for all other curves, we assume $1/\kappa_\phi = \SI{500}{\milli\second}$. (c) The participation ratios corresponding to the memory infidelities in (b) are plotted. The parameter values are displayed in Table~\ref{tab:parameters}. }\label{Fig:STORAGE}
\end{center}
\end{figure}

Finally, the total memory infidelity (consisting of write, idling, and read operations) is plotted in Fig.~\ref{Fig:SWAPIDLE}. This figure clearly shows the trade-off between fast control [attainable for large participation ratio $(g/\Delta)^2$] and slow cavity decoherence during idling (requiring small participation). For each experimental setup, the compromise between these two requirements gives rise to a different optimum in terms of intrinsic cavity lifetime and participation ratio. The best performance is achieved with the resonantly driven coupler with the minimum error given by equal swap and idling errors, $2\bar{\epsilon}_{\rm BS} = \bar{\epsilon}_i$. (The factor of 2 stems from the fact that we swap the state twice, first into, and then out of, the cavity.) For the idling time $t_i = \SI{1}{\milli\second}$ and intrinsic lifetime $1/\kappa = \SI{1}{\second}$, we find the optimum participation ratio to be $(g/\Delta)^2 \approx 4.5\times 10^{-4}$ and total error $\bar{\epsilon}_m \approx 3.7\times 10^{-3}$ which is in good agreement with numerical simulations [optimal participation ratio $(g/\Delta)^2 \approx 4\times 10^{-4}$ and error $\bar{\epsilon}_m \approx 3.8\times 10^{-3}$]. Given the much shorter swap times [$t_s\lesssim\SI{10}{\micro\second}$; see also Fig.~\ref{Fig:SWAP}(b)] compared to the idling time ($t_i = \SI{1}{\milli\second}$), the results are almost identical to having the total memory time $t_m = 2t_s+t_i = \SI{1}{\milli\second}$.

To gain more insight into the trade-off between control and storage, we study the memory infidelity as a function of the memory time $t_m$ with the SNAIL coupler in Fig.~\ref{Fig:STORAGE}. The shortest memory time is set by the speed of the beam-splitter operation, $t_{m,{\rm min}} = 2t_{\rm BS}$ for a given participation ratio. For intrinsic lifetime of $1/\kappa = \SI{10}{\milli\second}$ and $(g/\Delta)^2 = 0.1$ (the maximum participation for which the dispersive approximation holds), this gives $t_{m,{\rm min}} = \SI{103}{\nano\second}$. Already this shortest possible memory time outperforms idling in the transmon. This can be seen from comparing the total memory infidelity of the cavity $\bar{\epsilon}_m = 2\bar{\epsilon_{\rm BS}} = \frac{1}{3}\gamma t_{\rm BS}+\frac{1}{4}\gamma_\phi t_{\rm BS}$ with idling in the transmon, $\bar{\epsilon}_{i,{\rm tmon}} = \frac{1}{3}(2\gamma+\gamma_\phi)t_{\rm BS}$, which is independent from the intrinsic cavity lifetime.

\begin{figure}
\begin{center}
	\includegraphics[width=1.0\linewidth]{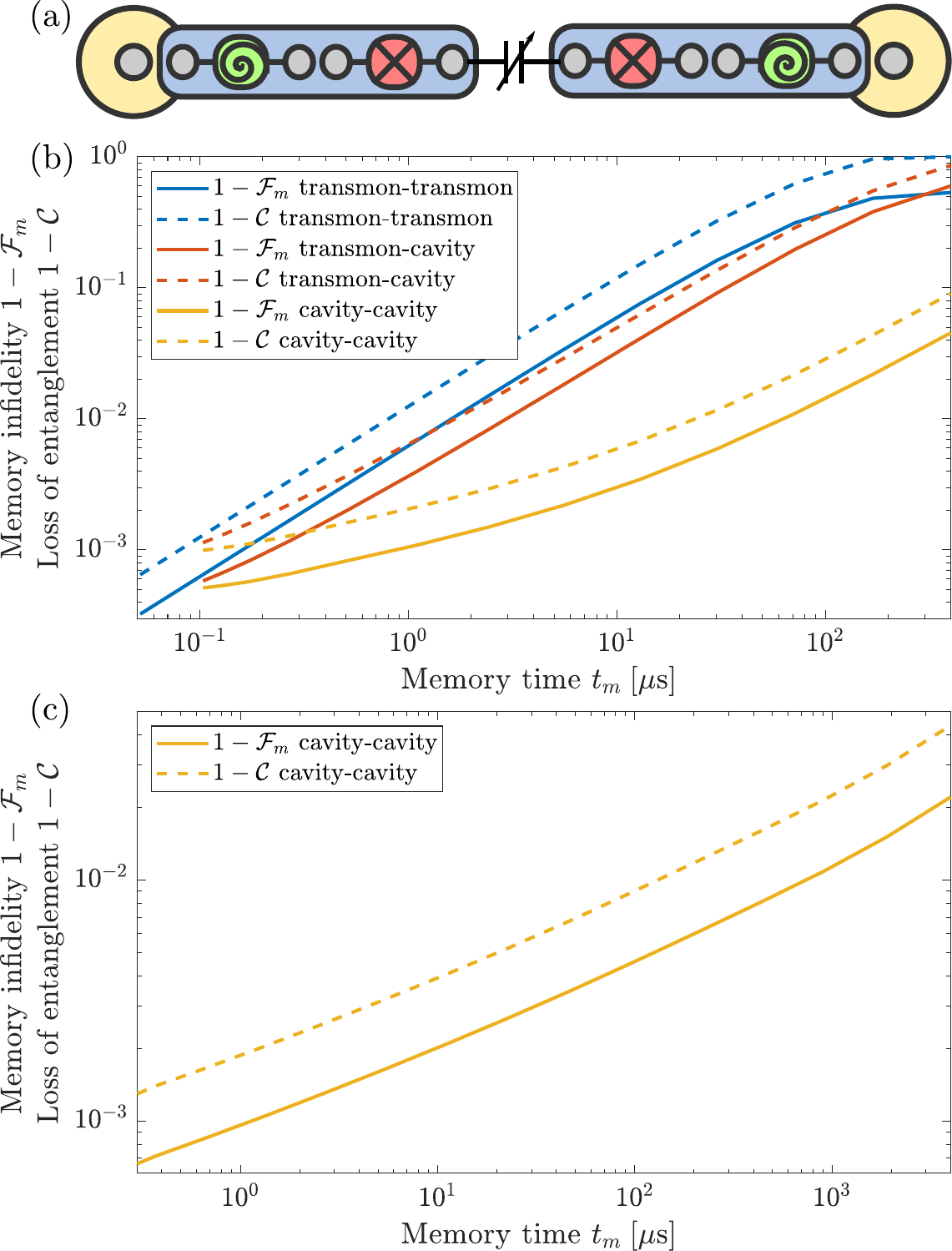}
	\caption{Preservation of the Bell state $\ket{\Phi_+} = (|00\rangle +|11\rangle)/\sqrt{2}$ during idling. (a) Schematic setup consisting of two transmons, each with a dedicated cavity memory (coupled via a SNAIL coupler). Tunable coupling between the transmons can be used to generate the Bell state. (b) Memory infidelity (solid lines) and loss of entanglement (quantified by concurrence, dashed) of the Bell state when it is left to idle in the transmons (blue lines), compared to when one of the transmon states is swapped to a cavity memory ($1/\kappa = \SI{10}{\milli\second}$) to idle and at the end swapped back to the transmon (red lines), and when both qubits are swapped to cavity memories and back (yellow lines). (c) Memory infidelity (solid line) and the loss of entanglement (dashed) of the Bell state when using a cavity memory with $1/\kappa = \SI{1}{\second}$ for both states. In both plots, the cavity participation ratio is optimized for each memory time $t_m$. The other parameter values are displayed in Table~\ref{tab:parameters}.}\label{Fig:BellState}
\end{center}
\end{figure}

Increasing the cavity lifetime to one second allows us to extend the memory time to the millisecond range as shown in Fig.~\ref{Fig:STORAGE}(b). When optimizing the participation ratio, we can reach error of 0.1 \% for $t_m = \SI{30}{\micro\second}$ and 1 \% for $t_m = \SI{6}{\milli\second}$. Using fixed participation ratio (we assume critical inverse Purcell decay, $\kappa = \kappa_\gamma$) leads to overall worse performance. Comparison of the optimal participation ratio with the critical coupling [panel (c)] shows that the memory fidelity is limited by the control for short memory times and by idling for long times. For short times, the critical participation ratio is below the optimum, resulting in slow beam-splitter gates between the transmon and cavity. On the other hand, the critical participation ratio is too large for long memory times, resulting in fast inverse Purcell decay during idling. The crossover between these two regimes occurs at about $t_m \approx \SI{3}{\milli\second}$.

Finally, to go beyond single-qubit storage, we numerically analyze preservation of entanglement in Fig.~\ref{Fig:BellState}. We assume an initial Bell state $\ket{\Phi_+} = (\ket{00}+\ket{11})/\sqrt{2}$ in two transmons with the possibility of swapping both qubits into separate cavity memories as depicted schematically in panel (a). Already swapping one qubit into a memory (intrinsic cavity lifetime $1/\kappa=\SI{10}{\milli\second}$) provides an advantage compared to keeping both states in transmons (as quantified by both the memory infidelity for the state $1-\mathcal{F}_m$ and loss of concurrence $1-\mathcal{C}$~\cite{Wootters1998}) {showing the viability of performing quantum gates on one qubit while idling the other as discussed in the introduction. However,} only swapping both qubits to memories provides a significant improvement. Similar to single-qubit memory (cf. Fig.~\ref{Fig:STORAGE}), swapping is advantageous {whenever the swap time is at most half the total memory time, $t_s\leq \frac{1}{2}t_m$}. With a longer cavity lifetime [$1/\kappa = \SI{1}{\second}$ in Fig.~\ref{Fig:BellState}({c})], the memory time can be increased similar to single-qubit states---we can push the memory time to $t_m = \SI{700}{\micro\second}$ while keeping the infidelity below 1 \%, which corresponds to single-qubit infidelity of 0.5 \%. The concurrence is generally slightly lower than the state fidelity but it keeps the same scaling with memory time.

\subsection{Consequences for quantum algorithms}

\label{ssec:algorithms}
\begin{figure*}
    \centering
    \includegraphics[width=\linewidth]{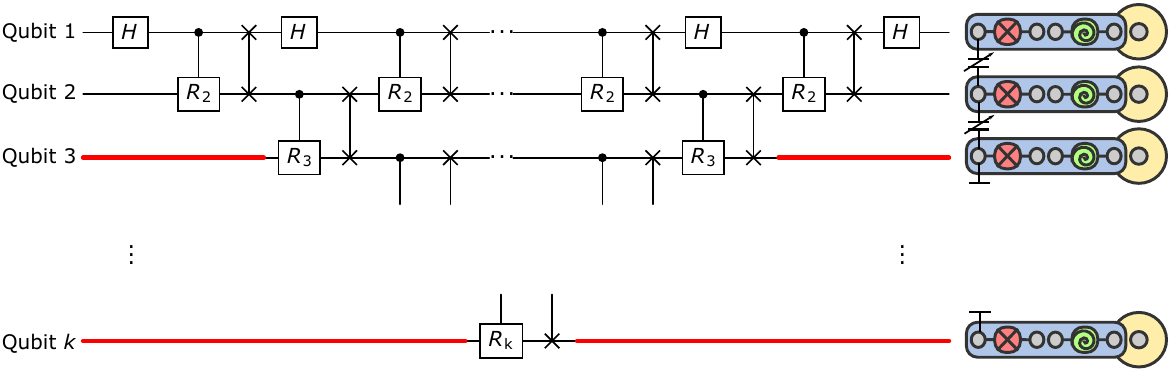}
    \caption{Quantum circuit for the quantum Fourier transform on a nearest-neighbor-coupled qubit array with $k$ qubits. Implementation requires Hadamard gates $H$, controlled qubit rotations $R_j = {\rm diag}[1, \exp(2\pi i/2^j)]$, and swaps. We assume that the algorithm is being implemented on a linear chain of transmon qubits with each qubit having a dedicated cavity memory (right), to which it is coupled via a SNAIL coupler and which is used to store the qubit state during idling. Two-qubit gates are implemented using tunable coupling between neighboring qubits. The thick red lines show the idling time spent in the cavities (which does not apply to the first two qubits in the array).}
    \label{fig:QFT}
\end{figure*}

To better understand the consequences of the improved information storage, we now consider the application of cavity memories for quantum Fourier transform (QFT) performed on a linear chain of transmon qubits, see Fig.~\ref{fig:QFT}. The quantum circuit for performing QFT on a chain of nearest-neighbor-coupled qubits (such as transmons) has been known for some time~\cite{Fowler2004} and shows quite a low gate density~\cite{QASMBENCH_10.1145/3550488}. As is apparent from the circuit diagram in Fig.~\ref{fig:QFT}, the gates are distributed unevenly among the qubits---the first qubit is almost never idle whereas the last qubit nearly always is. It is then natural to expect that this long idling can be improved by swapping the qubit state to a memory for better storage. In principle, even better performance could be achieved with all qubits primarily stored in cavity memories which would all be connected to only a few transmon qubits for performing gates. Such an architecture would allow for better qubit connectivity and would therefore reduce the number of gates needed to perform the QFT. However, designing a device in which a single transmon efficiently couples to a large number of microwave modes (whether in one or multiple cavities) may present considerable challenges. Alternatively, quantum acoustodynamical devices with high-overtone bulk acoustic resonators~\cite{Chu2018,Hann2019,Lupke2023} could present a suitable architecture if the lifetimes of acoustic modes are improved, but their detailed analysis would be beyond the scope of our work.

To evaluate possible fidelity improvements enabled by storing the qubit states in memories during idling, we now make several simplifying assumptions: First, we expect that the time needed to write into and read out of the memory is longer than the time needed to perform one- and two-qubit gates for the QFT algorithm (which is indeed the case for the control strategies described above). We therefore do not expect to store the first two qubits in their memories during the computation at all and only assume that qubits 3 to $k$ are stored at the beginning and end of the computation before and after they are needed (as highlighted by the thick red lines in Fig.~\ref{fig:QFT}). Second, we set all one- and two-qubit gate times to be equal, denoting them $t_g$, which allows us to simplify the total idling time for all qubits. Finally, we assume that all errors are small, $\bar{\epsilon}\ll 1$, and we can add them up linearly instead of multiplying all fidelities and evaluating the exponential decay of the stored quantum information. (This assumptions is key also for ensuring that the whole computation works with high fidelity.)

The total time that the qubit $j\geq 3$ spends idling before it is used in the computation is $t_j = (2j-3)t_g$, with equal time spent idling after the computation so the total idling time is $2t_j$. The total error qubit $j$ accumulates during idling (assuming the state is kept in the transmon) is therefore
\begin{equation}
    \bar{\epsilon}_{j,{\rm tmon}} = \frac{1}{3}(2\gamma+\gamma_\phi) t_j = \frac{2j-3}{3}(2\gamma+\gamma_\phi) t_g,
\end{equation}
where we assume relaxation and pure dephasing of the qubit during idling and average over all possible qubit states~\cite{Abad2022}.
The cumulative idling error can then be obtained by summing the individual idling errors,
\begin{equation}\label{eq:QFT_idle}
    \bar{\epsilon}_{m,{\rm tmon}} = \sum_{j=3}^k \bar{\epsilon}_{j,{\rm tmon}} = \frac{k(k-2)}{3}(2\gamma+\gamma_\phi) t_g.
\end{equation}
On the other hand, the decay rate in a cavity memory is much smaller, $\kappa_{\rm tot} = 2\Gamma_\downarrow + \Gamma_\phi = 2(\kappa+\kappa_\gamma)+\kappa_\phi+\kappa_{\gamma_\phi}\ll 2\gamma+\gamma_\phi$, but this improvement is partially offset by the errors associated with the write and read operations, $2\bar{\epsilon}_{\rm BS}$ (we assume the coupler scheme is used since it gives the best performance in Fig.~\ref{Fig:SWAPIDLE}). For each qubit, two write and read operations are needed and the total idling time in the memory is smaller than $2t_j$ by the time it takes to perform these operations, $4t_{\rm BS}$. The $j$th qubit memory error is therefore $\bar{\epsilon}_j = 4\bar{\epsilon}_{\rm BS} + \frac{1}{3}\kappa_{\rm tot}(t_j - 2t_{\rm BS})$ and the cumulative memory error is obtained by summing over all individual qubit errors,
\begin{equation}\label{eq:QFT_mem}
    \bar{\epsilon}_m = 4(k-2)\bar{\epsilon}_{\rm BS} + \frac{k-2}{3}\kappa_{\rm tot}(k t_g - 2t_{\rm BS}).
\end{equation}

\begin{figure}
\begin{center}
	\includegraphics[width=1.0\linewidth]{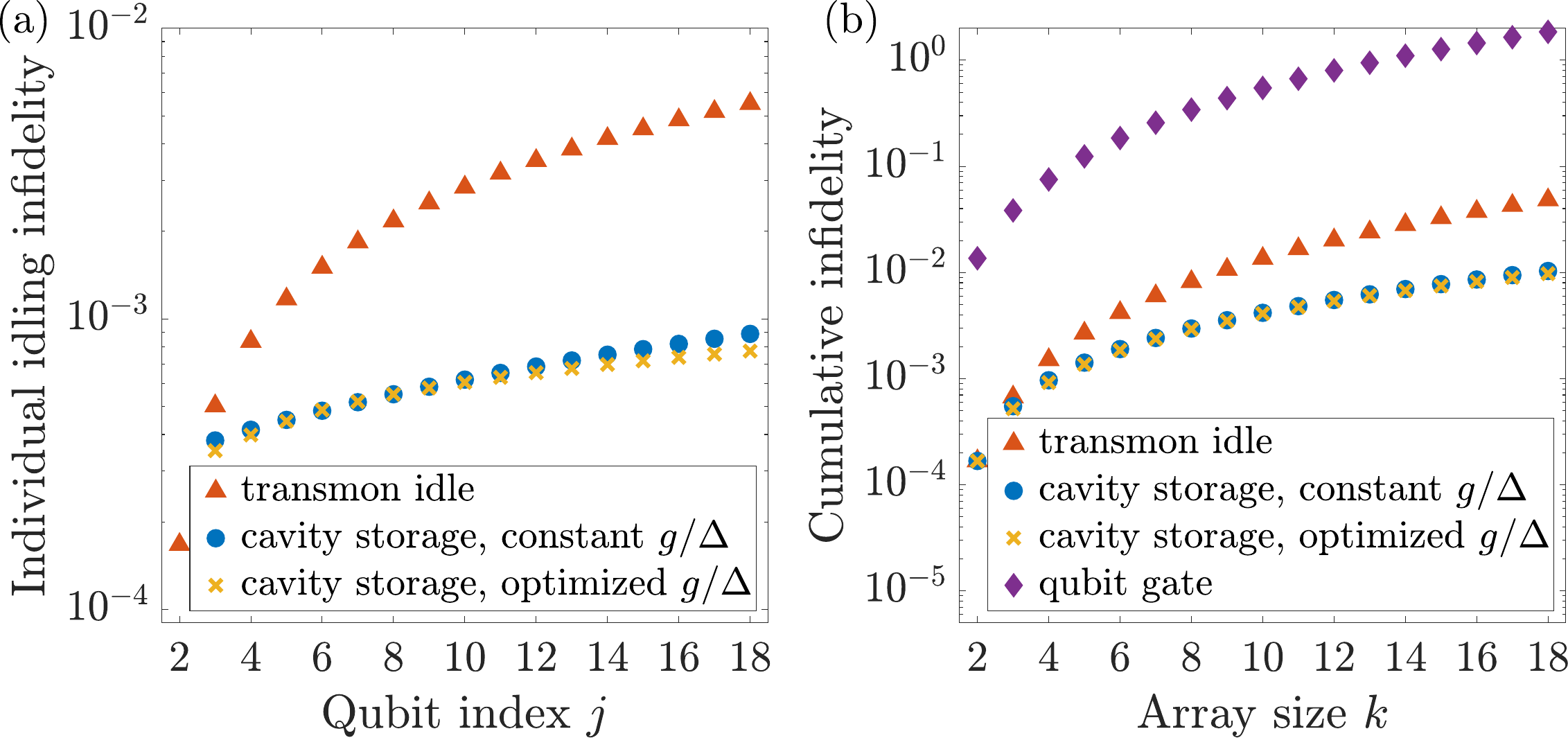}
	\caption{Simulations of average infidelities during QFT. (a) Individual qubit idling infidelities (total idling time $2t_j$) when idling in transmons (red triangles), in cavities with constant participation ratio (giving swap time $t_{\rm BS} = \SI{60}{\nano\second}$, blue circles), and in cavities with $(g/\Delta)^2$ optimized for each qubit separately (yellow crosses). (b) The cumulative infidelity in an array of $k$ qubits (obtained by summing individual infidelities for $j\leq k$). In addition to the idling infidelities [with the same symbols as in panel (a)], we also plot the cumulative gate infidelity (purple diamonds). The cumulative gate infidelity becomes larger than unity for large qubit arrays since we are adding individual gate errors. We use a storage cavity with a lifetime of $1/\kappa = 10$ ms with other parameter values taken from Table~\ref{tab:parameters}.}\label{Fig:QFTerror}
\end{center}
\end{figure}

To quantify the improvement, we assume a typical gate time $t_g = \SI{40}{\nano\second}$~\cite{Barends2014}. The longest swap time that still allows us to swap the third qubit into the cavity is $t_{\rm BS} = \SI{60}{\nano\second}$. From Fig.~\ref{Fig:SWAP}, the minimum participation ratio needed to achieve this swap time is $(g/\Delta)^2 = 0.07$ and the corresponding swap error $\bar{\epsilon}_{\rm BS} = 1\times 10^{-4}$. Setting $1/\kappa = \SI{10}{\milli\second}$ and taking the other values from Table~\ref{tab:parameters}, we get from Eq.~(\ref{eq:QFT_mem}) {the cumulative memory error}
\begin{equation}
    \bar{\epsilon}_m = (k-2)(3.6\times 10^{-4} +1.3\times 10^{-5} k) \,.
\end{equation}
On the other hand, assuming qubit parameters as in Table~\ref{tab:parameters} gives the cumulative idling error [cf. Eq.~\eqref{eq:QFT_idle}]
\begin{equation}
    \bar{\epsilon}_{m,{\rm tmon}} = 1.7\times 10^{-4}k(k-2) \,.
\end{equation}
For large arrays, the leading contribution is from the quadratic term which is smaller for cavity memories by a factor of 13. However, a slight improvement is possible already for $j=3$ ($\bar{\epsilon}_m = 4.0\times 10^{-4}$ for cavity memory and $\bar{\epsilon}_{m,{\rm tmon}} = 5.1\times 10^{-4}$ for idling in transmon); see also Fig.~\ref{Fig:QFTerror}. The {cumulative} memory error remains small ($\bar{\epsilon}_m<0.01$) for $k< 18$ compared to $k< 9$ for idling in transmons ($\bar{\epsilon}_{m,{\rm tmon}}<0.01$). Using state-of-the-art cavity memories can thus double the size of the qubit array.

The idling time varies significantly across the qubit array and choosing a single value of the participation ratio for each memory is generally not optimal [cf. Fig.~\ref{Fig:STORAGE}(b)]. Optimizing the participation ratio for each qubit separately gives improved memory fidelity as can be seen in Fig.~\ref{Fig:QFTerror}. The participation ratio $(g/\Delta)^2 = 0.07$ is near-optimal for $j = 6,7$ but the individual memory errors can be improved for all the remaining qubits. For $j=3$, it is better to speed up the swap (by increasing the participation ratio) from the maximum duration $t_{\rm BS} = \SI{60}{\nano\second}$ and have nonzero idling time in the memory. For large $j$, it is advantageous to use a smaller participation ratio with slower swap but better idling due to reduced inverse Purcell decay. However, the difference between constant and optimized participation ratio remains relatively small for the cumulative memory errors, especially when compared with the cumulative gate error (discussed below).

To place this improvement into proper context, it is also crucial to estimate the total error due to gates. The circuit uses $k$ single-qubit Hadamard gates and $k(k-1)$ two-qubit gates (controlled rotations and swaps). Assuming state-of-the-art performance with single-qubit gate error $\bar{\epsilon}_{1q} = 8\times 10^{-4}$ and two-qubit gate error $\bar{\epsilon}_{2q} = 6\times 10^{-3}$~\cite{Barends2014} (and neglecting the idling error in the first two qubits, which is much smaller than the gate error), we get the {cumulative} gate error
\begin{equation}\label{eq:QFT_qubit}
    \bar{\epsilon}_{q} = 6k^2 \times 10^{-3} - 5.2k\times 10^{-3}.
\end{equation}
The gate error is therefore larger than the idling error (for both cavity memories and transmons) by an order of magnitude, as can also be seen in Fig.~\ref{Fig:QFTerror}(b). Already the single-qubit gate error is an order of magnitude larger than the corresponding transmon idling error with the two-qubit gate error larger by another factor of 7.5. With future improvements in gate design, the idling error may become more prominent---improving the average gate error to $10^{-4}$ per qubit (such that $\bar{\epsilon}_{1q} = 10^{-4}$ and $\bar{\epsilon}_{2q} = 2\times 10^{-4}$) would result in the idling error (without cavity memories) comparable to the gate error. This can be easily seen from the individual errors---the individual idling error in transmon qubits is $\bar{\epsilon}_{\rm tmon} = 8\times 10^{-5}$ per gate. Given that the gate density is about half the product of the circuit width and depth, all errors being equal, idling errors become as important as gate errors.

We have not computed the fidelity of the noisy QFT algorithm averaged over all possible input states.  However, our results on gate error probabilities can be used to obtain a lower bound on the the algorithmic fidelity.  If we make the pessimistic assumption that any gate error results in zero overlap between the actual and ideal output state, then the probability of zero errors occurring sets a lower bound on the algorithmic fidelity.  This assumption has been shown to provide a good estimate of the fidelity in Google's random circuit experiments \cite{arute_quantum_2019} but its validity for highly structured algorithms (such as the QFT) might be questionable (i.e., too pessimistic).  One additional caveat is that even if there are zero errors (i.e., no quantum jumps) it is possible that the no-jump evolution is not proportional to the identity operation.  This is not the case for qubit Pauli errors, but is the case for amplitude damping of qubits and cavities. This could then potentially increase the algorithm fidelity beyond the errors calculated here, but from the requirement that the Kraus map evolution is trace preserving,  it follows that the errors due to no-jump evolution cannot be significantly worse than the errors from the jumps.  Hence  the probability of zero (jump) errors can be expected to be close to the infidelity of the QFT circuit.

\section{Discussion}

\subsection{Cavity reset}

In addition to efficient control and readout of the cavity state, it is important to be able to quickly reset ultra-long-lived cavities to a known fiducial state (e.g., Fock state $\ket{0}$).  This could be achieved algorithmically by measurement (say by photon number measurement to collapse the cavity to a Fock state) and then applying appropriate gates to map the measured Fock state onto vacuum. High-quality reset is essential for subsequent high-fidelity state preparation and protocols involving repeated measurements have been shown to be very effective for this purpose in multi-level qubit systems \cite{PhysRevA.98.022305,ElderPhysRevX.10.011001}. However, such algorithmic reset works only in the strong dispersive regime which allows photon-number splitting of the qubit spectrum. For ultra-high-$Q$ cavities, this requirement introduces strong inverse Purcell decay and is therefore unsuitable. An alternative would be to devise reset based on measurements of the characteristic function which is possible in the weak dispersive coupling thanks to a strong cavity displacement~\cite{Campagne2020}.

A better strategy might therefore be to reset the cavity in analog fashion by temporarily $Q$ \emph{switching} the cavity using a nonlinear element. With suitable driving, this auxiliary circuit can act as a frequency-converting beam-splitter to rapidly and irreversibly dump the cavity contents into a low-$Q$ cold mode \cite{PhysRevLett.114.090503,Sivak_GKP_2022}. This would be possible without additional hardware by using the qubit to drive a beam-splitter interaction between the cavity and the qubit's readout resonator. A potential limitation of this approach is that it can only cool the cavity to the same thermal level as experienced by the low-$Q$ element which, in the case of a transmon, might be higher than the thermal occupation of the cavity bath.

\subsection{Standard quantum limit for memory control}

As we have demonstrated, using microwave cavity modes as quantum memories requires a compromise between fast, efficient control and long storage time. A rough but useful analogy to gain intuition about this trade-off is that of the standard quantum limit, commonly applied for position measurements and force sensing in cavity optomechanics and electromechanics~\cite{Clerk2010,Aspelmeyer2014}. In these position measurements, the optimal sensitivity can be reached for probe fields that balance the measurement imprecision and backaction noise. Imprecision noise, the finite variance of the quadrature onto which information about the mechanical position is imprinted, masks the signal while backaction noise, caused by random momentum kicks to the mechanical oscillator by incoming photons, affects the evolution of the mechanical mode, reducing the sensitivity of future measurements. For coherent (semiclassical) states, imprecision noise can be reduced by increasing the photon number (scaling as $1/n$), but this also increases the backaction noise (directly proportional to $n$); the intensity at which both noise contributions are equal defines the standard quantum limit.

Although the underlying physical mechanism for controlling microwave cavity memories is different, similar effects are at play. For control schemes based on strong dispersive coupling (such as gradient ascent pulse engineering~\cite{Heeres2017} or SNAP gates~\cite{Heeres2015,Kudra2022}), the competition between control and storage has the same form as expressed in the memory infidelity (refer back to Sec.~\ref{sec:control} for derivation)
\begin{equation}
    1 - \mathcal{F}_m = \frac{\alpha\gamma_{\rm tot}}{\chi} + \frac{{\gamma_{\rm tot}}\chi}{2K}nt_i + \kappa_{\rm tot}nt_i.
\end{equation}
The imprecision in cavity control (first term on the right-hand side) can be reduced by increasing the dispersive coupling strength $\chi$ between the qubit and cavity but this increases the backaction of the control by enhancing the inverse Purcell decay of the cavity mode (second term on the right-hand side). The optimal memory fidelity is reached when both terms are equal, corresponding to the standard quantum limit for memory control. The term $\kappa_{\rm tot}nt_i$ {describing intrinsic cavity decoherence} is independent of $\chi$ and plays the role of thermal noise in optomechanical measurements.

The value of the optimal coupling in general depends on the expected {total storage} time $t_i$ with longer times requiring weaker coupling. For a given coupling rate $\chi$, the fidelity of long storage is limited by backaction from the inverse Purcell decay and the fidelity of short storage is limited by imprecision in control. This behavior is again analogous to the standard quantum limit in optomechanics where the optimal probe strength depends on the frequency of the signal of interest. Detection of low-frequency signals (associated with long times) is then limited by measurement backaction and detection of high-frequency signals (corresponding to short times) by imprecision~\cite{Aspelmeyer2014}.

In optomechanics, the standard quantum limit does not represent an ultimate limit. In optomechanics, measurement backaction can be avoided by performing single-quadrature measurements~\cite{Hertzberg2009,Suh2014,Shomroni2019}, modifying the detection~\cite{Ockeloen-Korppi2018,Mason2019}, or engineering negative-mass reference frames~\cite{Moller2017,MercierDeLepinay2021}. Since the physics underpinning the standard quantum limit for memory control is different, these techniques cannot be applied, but the standard quantum limit can still be surpassed. Interestingly, for microwave cavity memories, the backaction is the fundamental and unavoidable limitation while imprecision can be reduced by suitable techniques. Using conditional displacements~\cite{Eickbusch2021X, diringer2023X}, sideband driving~\cite{Murch2012}, and three- or four-wave mixing~\cite{RosenblumCavity2023,Ben_Stijn_preprint} allows the scaling of the control error to be improved from $1/\chi$ to $1/\sqrt{\chi}$ thanks to strong cavity, transmon, and coupler driving. Using a buffer cavity allows the control fidelity to be improved even further, at least in principle, as the figure of merit determining the speed of transferring the state from/to the qubit is the participation ratio between the qubit and buffer. With a whole chain of such buffer cavities, a large gap between the lifetimes of the qubit and storage cavity can be bridged using smaller steps between the (inverse-Purcell-limited) lifetimes of these buffer cavities.

\subsection{Limitations on long-time storage}

Based on our analytical models of channel fidelities, we can estimate the maximum storage times attainable with state-of-the-art nonlinear circuits. These limits are summarized in Table~\ref{tab:limits} for the SNAIL coupler and cavity cascading. For high-fidelity (one percent error) swap and storage, memory time up to \SI{11}{\milli\second} is possible with the coupler; we assume critical coupling, $\kappa = \kappa_\gamma$ and optimize the participation ratio $(g/\Delta)^2$ {(which therefore determines also the necessary minimum intrinsic lifetime also shown in the table)}. At this point, the fidelity is limited primarily by intrinsic cavity dephasing (we assume pure dephasing lifetime $1/\kappa_\phi = \SI{500}{\milli\second}$~\cite{RosenblumCavity2023}). Unfortunately, little is known about the sources of this dephasing and how it might scale when improving the quality factor of the cavity. If we assume that intrinsic cavity dephasing can be removed in the future (either through understanding its origins or due to favorable scaling with the intrinsic quality factor), the achievable storage time increases to almost \SI{50}{\milli\second} which is 250 times longer than the assumed transmon relaxation.

\begin{table}
    \centering
    \caption{Maximum attainable memory times (with required intrinsic cavity lifetime) with SNAIL coupler and cavity cascading for target memory error $\bar{\epsilon}_m$. For both approaches, the participation ratios are optimized based on analytical formulas listed in Sec.~\ref{sec:memory_sim}. We assume the parameters given in Table~\ref{tab:parameters}; we also include results without intrinsic cavity dephasing, $\kappa_\phi = 0$. For the $e$--$f$ rotation error (for cavity cascading), we assume $\bar{\epsilon}_{ef} = 10^{-3}$ except for the smallest memory error, which becomes unattainable and for which we instead set $\bar{\epsilon}_{ef} = 0$.}
    \label{tab:limits}
\begin{tabular}{|l|ccc|}
    \hline
    Target memory error $\bar{\epsilon}_m$ & $0.001$ & $0.005$ & $0.01$ \\
    \hline \hline
    Coupler &&& \\
    Maximum memory time $t_{m,{\rm max}}$ & \SI{47}{\micro\second} & \SI{3}{\milli\second} & \SI{11}{\milli\second} \\
    Minimum intrinsic lifetime $1/\kappa$ & \SI{95}{\milli\second} & \SI{1.5}{\second} & \SI{3.6}{\second} \\
    \hline 
    Coupler ($\kappa_\phi = 0$) &&& \\
    Maximum memory time $t_{m,{\rm max}}$ & \SI{49}{\micro\second} & \SI{6}{\milli\second} & \SI{48}{\milli\second} \\
    Minimum intrinsic lifetime $1/\kappa$ & \SI{97}{\milli\second} & \SI{2.4}{\second} & \SI{9.6}{\second} \\
    \hline \hline 
    Cascading &&& \\
    Maximum memory time $t_{m,{\rm max}}$ & \SI{252}{\micro\second} & \SI{4.2}{\milli\second} & \SI{17}{\milli\second} \\
    Minimum intrinsic lifetime $1/\kappa$ & \SI{1.1}{\second} & \SI{8.9}{\second} & \SI{26.9}{\second} \\
    \hline
    Cascading ($\kappa_\phi = 0$) &&& \\
    Maximum memory time $t_{m,{\rm max}}$ & \SI{413}{\micro\second} & \SI{102}{\milli\second} & \SI{10}{\second} \\
    Minimum intrinsic lifetime $1/\kappa$ & \SI{1.6}{\second} & \SI{126}{\second} & \SI{5.2e3}{\second} \\
    \hline
\end{tabular}
\end{table}

The storage time can be greatly extended with cavity cascading. Assuming again realistic parameters (with the same intrinsic cavity dephasing), we expect storage times up to \SI{17}{\milli\second} with 99 \% fidelity {({which, due to the assumed} {intrinsic dephasing}, is not much larger than with the SNAIL coupler)}. Removing cavity dephasing improves this bound to almost 10 seconds which is nearly five orders of magnitude longer than the lifetime of transmons and SNAILs. {When demanding smaller total error, an even larger advantage of cavity cascading can be seen even with intrinsic dephasing---for target error $\bar{\epsilon}_m\leq 10^{-3}$, cavity cascading allows storage times five times longer than using the SNAIL coupler (albeit with cavity lifetime larger by an order of magnitude).} {These improvements demonstrate} the viability of buffer cavities as a tool to bridge the gap between ultra-long-lived microwave cavities and noisy nonlinear circuits for their control. Moreover, these results highlight the importance of understanding the sources of cavity dephasing and its scaling when improving the lifetime (through surface treatments, material choice, or mode volume and geometry). These results also suggest the possibility of using a multi-cavity repetition code (such as has been proposed for cat qubits~\cite{Regent2023}) if it turns out that the error channel is indeed heavily biased towards dephasing relative to amplitude damping in ultra-high $Q$ cavities.

\subsection{Cavity loss beyond inverse Purcell decay}\label{sec:losses}

Throughout this paper, we have assumed that the lifetime of the storage cavity is limited by the inverse Purcell effect (photon decay due to  absorption of the  qubit).  The validity of this assumption is unclear and requires further experimental investigation. There is experimental evidence that additional losses associated with the dielectric material of the qubit chip present a stronger decay channel than the inverse Purcell decay \cite{RosenblumCavity2023,Read2022DielectricDipper}. Although the exact mechanisms for these losses are unclear, a simple estimate of the dielectric losses and their effect on the memory fidelity can be obtained as follows: We scale the inverse Purcell decay rate by a phenomenological constant $R$ that takes the dielectric loss of the cavity in the qubit substrate into account,
\begin{equation}
    \kappa_\gamma = R\gamma\left(\frac{g}{\Delta}\right)^2.
\end{equation}
This expression cannot fully account for all loss channels and their scaling with system parameters since the dielectric loss {is not expected to depend strongly} on the detuning between the qubit and cavity; however, assuming a fixed detuning and a variable insertion depth of the qubit chip into the cavity, both qubit coupling $g$ and dielectric loss would be expected to increase or decrease together, meaning that $R$ is reasonably well-defined in this case. 

To estimate the size of the constant $R$, we can use the recent experiment by Milul \emph{et al}.\ \cite{RosenblumCavity2023}. The bare cavity (without qubit chip) has a relaxation lifetime $T_0 = 1/\kappa = \SI{30}{\milli\second}$. After inserting the qubit chip, this value drops to $T = 1/(\kappa+\kappa_\gamma) = \SI{25.6}{\milli\second}$, suggesting inverse Purcell lifetime $T_\gamma = 1/\kappa_\gamma = T_0T/(T_0-T) = \SI{175}{\milli\second}$. This observed lifetime is smaller than the inverse Purcell effect (for the particular detuning used in this experiment) estimated by the authors ($T_\gamma\approx$ \SI{278}{\milli\second}) by a factor $R= 1.6$. We therefore estimate that $R$ moderately larger than unity ($1 < R\lesssim 2$) can account for all loss channels associated with the qubit chip in near-future devices.

Going forward, an important question is how these additional cavity loss channels scale when improving the $Q$ factor. The very highest-$Q$ cavities are achieved not only by control of the surface chemistry but also by making the cavity larger (which increases the $Q$ factor without changing the finesse).  Large cavities have resonances more closely spaced in frequency and multi-mode effects (which are known to be important to the Purcell effect \cite{MultimodePurcell_PhysRevLett.101.080502}) may worsen the inverse Purcell rate, making the phenomenological factor $R$ larger for these large, ultra-high-$Q$ cavities with smaller free spectral range. 
Another risk associated with cavities with a large number of modes is dephasing of the qubit due to dispersive coupling to thermal photons occupying these modes \cite{PhysRevB.86.180504}. This can be especially problematic if any of the higher resonance frequencies lie outside the range of filters on the input lines.  Additionally, higher frequency resonances may have shorter lifetimes making it more difficult to use echo techniques to protect the qubit against thermal photon induced dephasing.
Deeper understanding of cavity losses, in particular dielectric loss in the qubit substrate, scattering into nearby cavity modes, as well as identifying any additional loss channels, is therefore crucial for developing ultra-high-$Q$ cavity modes for quantum technologies.

\section{Conclusions}

In summary, we have investigated the performance of ultra-high-quality microwave cavities as quantum memories for superconducting qubits. We derived the dissipators stemming from hybridization of the cavities with nonlinear circuits used for their control and developed analytical models capturing the main errors in swapping quantum states to/from the cavity and during idling. We were thus able to evaluate the trade-offs between fast control and long-term storage and determine memory-time limits for current nonlinear devices and their main limitations. Although we focused only on passive storage of a qubit state in the cavity (with the only active control being the write and read operations with the transmon), similar trade-offs can be expected when applying quantum gates directly on the Fock-encoded qubit in the cavity.

Good protection of cavity modes against transmon-induced decoherence requires suppressing cross-Kerr interaction during idling, as this interaction mediates the propagation of dephasing to the cavity when transmon heating occurs. This suppression can be achieved through the use of a suitable coupler such as a SNAIL~\cite{SivakSNAIL,Ben_Stijn_preprint} which can be operated at the Kerr-free point. Such a setup does not protect from the inverse Purcell decay, which will require more sophisticated approaches based on large on-off ratio couplers~\cite{zhao2023integrating,Noh2023}. Additionally, strategies inspired by qubit cloaking might offer protection against qubit-induced cavity decoherence~\cite{Lledo2023}. Finally we note that in addition to tunable couplers activated by microwave driving, there exist couplers tunable via magnetic flux that have been used for qubit-qubit coupling \cite{tunablecouplerMartinis2014,tunablecoupler_Oliver}, and these may also be good candidates for control of ultra-high-\textit{Q} cavities.

Further improvements to gate fidelity can be made with the use of fault-tolerant schemes using three or more levels in the auxiliary qubit. Typically, these schemes assume the logical auxiliary qubit in the $g$--$f$ manifold with the $\ket{e}$ state serving as a flag to detect qubit decay. For the sideband transitions we considered here, distinguishing between the transmon being in the ground and first excited state tells us whether the transition succeeded (for ground state), allowing us to conditionally improve swap fidelity through post-selection. More sophisticated control schemes exist where detection of the flag state does not introduce backaction and the gate can be repeated until success~\cite{Reinhold2020ErrorCorrectedGate,PathIndependent_PhysRevLett.125.110503,Ma2022} with similar strategies existing for fault-tolerant measurements~\cite{ElderPhysRevX.10.011001,teoh2022dualrail}.

Going forward, more refined models will need to determine the losses associated with the superconducting circuit chips where dielectric bulk and surface losses seem to pose more stringent limitations than those based on inverse Purcell decay through the qubit. New package geometries that reduce the amount of chip substrate exposed to the cavity may help with this issue. Similar efforts will have to focus on modelling and estimating intrinsic cavity dephasing. State-of-the-art cavities have intrinsic dephasing lifetimes over half a second~\cite{RosenblumCavity2023}. However, the sources of this dephasing are unclear so it remains an open question how this dephasing changes when improving cavity relaxation through better surface treatments, using novel materials, or increasing the mode volume. With strong dephasing induced by nonlinear control circuits, it remains difficult to measure the intrinsic dephasing which currently hinders all such efforts. 

In our work, we focused on passively storing a qubit state in the vacuum and single-photon state of the resonator. However, one of the key advantages of the infinite-dimensional Hilbert spaces of linear resonators is the ability to efficiently encode quantum information in complex multi-photon states and correct errors accumulated during idling. On one hand, this opens the way to additional control techniques unsuitable for state swap between a transmon and cavity. A good example are gates based on conditional displacements~\cite{Eickbusch2021X,diringer2023X} where the (possibly weak) dispersive coupling is enhanced by a large intracavity photon number; since this is limited by the critical photon number, the control rate scales with the square root of the cavity participation ratio $\sqrt{(g/\Delta)^2}$. On the other hand, these states are more sensitive to relaxation (due to their higher photon numbers) and to dephasing through the cavity self-Kerr term. These considerations make such an analysis an appealing target for future work, particularly for evaluating the optimal duty cycle of error-correcting rounds and the consequences for surpassing the break-even point for quantum memories and computing with bosonic qubits.

\begin{acknowledgments}
    We thank Serge Rosenblum for a discussion on substrate losses. I.P. and O.\v{C}. thank projects JG\textunderscore{}2023\textunderscore{}001 of the Palack\'{y} University and LUAUS23012 of the Ministry of Education, Youth and Sports (MEYS \v{C}R). RF acknowledges grant 21-13265X of the Czech Science Foundation. I.P., O.\v{C}., and R.F. have been further supported by the European Union’s 2020 research and innovation programme (CSA—Coordination and support action, H2020-WIDESPREAD-2020-5) under Grant Agreement No. 951737 (NONGAUSS). A.M., J.W.O.G., and S.M.G. were supported by the U.S. Army Research Office (ARO) under grant W911NF-23-1-0051. A.E. was supported by, and S.M.G. acknowledges the additional support of, the U.S. Department of Energy, Office of Science, National Quantum Information Science Research Centers, Co-design Center for Quantum Advantage (C2QA) under contract No. DE-SC0012704. The views and conclusions contained in this document are those of the authors and should not be interpreted as representing the official policies, either expressed or implied, of the Army Research Office (ARO), or the U.S. Government. 
The U.S. Government is authorized to reproduce and distribute reprints for Government purposes 
notwithstanding any copyright notation herein.  

External interest disclosure: SMG receives consulting fees and is an equity holder in Quantum Circuits, Inc.
\end{acknowledgments}

\appendix
\section{Gaussian operations using three- and four-wave mixing}\label{sec:mixing}

Gaussian operations are those operations on (linear) cavity modes that map each Gaussian state onto a Gaussian state; Gaussian states are fully characterized in phase space by the mean values and covariance matrix of the quadrature operators $x = (a+a^\dagger)/\sqrt{2}$, $p=-i(a-a^\dagger)/\sqrt{2}$~\cite{Weedbrook2012}. Among these operations, phase space displacements and rotations are the simplest since they can be performed without requiring a nonlinear element. Phase space displacements can be achieved using a strong classical pump injected into the cavity through a weakly coupled port. Phase space rotations of the bosonic mode can be easily implemented \emph{in software} by modifying the phase of the microwave pulses applied to the cavity during subsequent operations. Other Gaussian operations (single- and two-mode squeezing and beam-splitter transformations) can be implemented using three- and four-wave mixing (although two-mode squeezing can be efficiently synthesized from single-mode squeezing plus beam-splitter operations).

To describe general three- and four-wave mixing processes, we consider the initial Hamiltonian
\begin{subequations}
\begin{align}
{H} &= {H}_c +{H}_q +{H}_{\rm dr}, \\
\begin{split}
{H}_c &= \omega_{a} {a}^\dagger{a} +\omega_{b} {b}^\dagger{b} +g_a({a}^\dagger{q} +{q}^\dagger{a}) +g_b({b}^\dagger{q} +{q}^\dagger{b}),
\end{split} \\
{H}_q &= \omega_{q} {q}^\dagger{q} +g_3({q}^\dagger +{q})^3 +g_4({q}^\dagger +{q})^4, \\ 
{H}_{\rm dr} &= \sum_k (e^{-i\omega_k t}\epsilon_k{q}^\dagger +e^{i\omega_k t}\epsilon_k^*{q}) \,,
\end{align}
\end{subequations}
where ${H}_c$ describes the two cavity modes with annihilation operators ${a},{b}$ and frequencies $\omega_{a,b}$, respectively. The coefficients $g_a$, $g_b$ describe the fields' coupling to the qubit with annihilation operator ${q}$. The Hamiltonian ${H}_q$ describes the qubit with frequency $\omega_{q}$ and third- and fourth-order nonlinearities $g_{3,4}$. The Hamiltonian ${H}_{\rm dr}$ contains the pumps that are used to drive the qubit. This is the most general form of the wave-mixing Hamiltonian we consider; single-mode control can be recovered by setting $g_b = 0$, transmon by $g_3 = 0$, and Kerr-free SNAIL by $g_4 = 0$.

We can derive the three- and four-wave mixing interactions by performing a sequence of transformations that take into account the dressing of the modes by their mutual interactions and drives and moving to a suitable rotating frame. The operators are dressed first by the interactions and then by the drives. Moving to a rotating frame in which all free oscillations vanish, we obtain the Hamiltonian~\cite{Grimm2020, Pietikainen2022}
\begin{equation}
\begin{split}
{H}_0 &= -\frac{K}{2}{q}^{\dagger 2}{q}^2-K_a a^{\dagger 2} a^2-K_b b^{\dagger 2} b^2 \\
&\quad -\chi_{a}{a}^\dagger{a}{q}^\dagger{q} -\chi_{b}{b}^\dagger{b}{q}^\dagger{q} -\chi_{ab}a^\dagger ab^\dagger b \\
&\quad -\chi_a'a^{\dagger 2}a^2q^\dagger q - \chi_b'b^{\dagger 2}b^2q^\dagger q,
\end{split}
\end{equation}
where $K = -12g_4, K_{a,b} = \frac{1}{2}K(g_{a,b}/\Delta_{a,b})^4, \chi_{a,b} = 2K(g_{a,b}/\Delta_{a,b})^2$, $\chi_{ab} = 2K(g_a/\Delta_a)^2(g_b/\Delta_b)^2$, $\chi_{a,b}'=2K(g_{a,b}/\Delta_{a,b})^4$, and $\Delta_{a,b} = \omega_{a,b}-\omega_q$. This is the Hamiltonian we have regardless of the driving frequencies. In addition to $H_0$, the resonances between the drives and the system's frequencies can create additional three-wave (cubic nonlinearity) and four-wave (quartic) mixing terms through the hybridized qubit mode $q + (g_a/\Delta_a)a + (g_b/\Delta_b)b +{\rm H.c.}$

Single-mode squeezing can be generated with three-wave mixing and a single drive tone at frequency $\omega_1 = 2\omega_a$, giving the Hamiltonian
\begin{equation}
{H}_{\rm SMS} = 3g_3\xi_1\left(\frac{g_a}{\Delta_a}\right)^2(a^2 + {a}^{\dagger 2}),
\end{equation}
where we assume $\xi_1=\epsilon_1/(\omega_1-\omega_q)\in\mathbb{R}$ for simplicity. Alternatively, we can use four-wave mixing with two drives of frequencies satisfying $\omega_1 +\omega_2 = 2\omega_a$,
\begin{equation}
{H}_{\rm SMS} = K\xi_1\xi_2\left(\frac{g_a}{\Delta_a}\right)^2(a^2 + {a}^{\dagger 2}) \,.
\end{equation}
Beam-splitter interaction between the cavity fields is created similarly either by three-wave mixing with pumping at the frequency difference of the two modes ($\omega_1 = \omega_a -\omega_b$),
\begin{equation}
{H}_{\rm BS} = 6g_3\xi_1\frac{g_a}{\Delta_a}\frac{g_b}{\Delta_b}({a}^{\dagger}{b} + {b}^\dagger{a}) \,,
\end{equation}
or by four-wave mixing with two pump frequencies $\omega_1 \pm\omega_2 = \omega_a -\omega_b$
\begin{equation}
{H}_{\rm BS} = 2K\xi_1\xi_2\frac{g_a}{\Delta_a}\frac{g_b}{\Delta_b}({a}^{\dagger}{b} + {b}^\dagger{a}) \,,
\end{equation}
Finally, two-mode squeezing can be engineered by three-wave mixing with pump frequency $\omega_1 = \omega_a+\omega_b$,
\begin{equation}
    H_{\rm TMS} = 6g_3\xi_1\frac{g_a}{\Delta_a}\frac{g_b}{\Delta_b}(ab+a^\dagger b^\dagger),
\end{equation}
or by four-wave mixing with pump tones $\omega_1\pm\omega_2 = \omega_a+\omega_b$,
\begin{equation}
    H_{\rm TMS} = 2K\xi_1\xi_2\frac{g_a}{\Delta_a}\frac{g_b}{\Delta_b}(ab+a^\dagger b^\dagger).
\end{equation}

An important insight is that the quality of single-mode squeezing does not depend on the lifetime of the mode and the quality of two-mode operations depends only on the ratio of their lifetimes. For efficient generation of single-mode squeezing, we demand that the squeezing is generated faster than the cavity decays, $g_{\rm SMS}/\kappa_{\rm tot}\gg 1$. For a critically coupled cavity mode, $\kappa_\gamma = \kappa = \frac{1}{2}\kappa_{\rm tot}$, this condition can be simplified to (assuming a SNAIL mixer)
\begin{equation}\label{eq:g_sq1}
    \frac{3g_3\xi_1}{\gamma} \gg 1,
\end{equation}
which is independent of $\kappa$ and $g/\Delta$; for a transmon mixer, we get the analogous condition $K\xi_1\xi_2/\gamma\gg 1$. For two-mode operations (beam-splitter and two-mode squeezing interactions), we get similar conditions comparing the strength of the coupling to the total decay rate of the faster decaying cavity mode. Assuming without loss of generality $\kappa_a > \kappa_b$ and critically coupled cavity modes, we can replace $g_j/\Delta_j = \sqrt{\kappa_j/\gamma}$ ($j = a,b$) and the condition $g_{\rm BS}/\kappa_{a,{\rm tot}}\gg 1$ for a SNAIL coupler becomes
\begin{equation}
    \frac{3g_3\xi_1}{\gamma} \gg \sqrt{\frac{\kappa_a}{\kappa_b}}
\end{equation}
with a similar condition existing for a four-wave mixer.

\section{Derivation of dressed master equation}\label{sec:masterEq}

To derive the master equation in the dispersive frame, we follow the approach of Ref.~\cite{Boissonneault2009} and apply the dispersive transformation
\begin{equation}
    U = \exp\left[\frac{g}{\Delta}(q^\dagger a-a^\dagger q)\right]
\end{equation}
on the system Hamiltonian and the system-bath interaction. Transforming the system Hamiltonian gives the standard cross-Kerr interaction,
\begin{equation}
    H_0 = -\frac{K}{2}q^{\dagger 2}q^2 -K_aa^{\dagger 2}a^2 - \chi a^\dagger aq^\dagger q - \chi'a^{\dagger 2}a^2 q^\dagger q,
\end{equation}
where $K_a = \frac{1}{2}K(g/\Delta)^4$, $\chi = 2K(g/\Delta)^2$, and $\chi' = 2K(g/\Delta)^4$. To derive the dressed dissipators, we start from the initial system-bath interaction of the form
\begin{equation}
\begin{split}
    H_{\rm SB} &= i\int d\omega [g_\kappa(\omega)a^\dagger b_\kappa(\omega) - g_\kappa^\ast(\omega)b_\kappa^\dagger(\omega) a] \\
    &\quad +i\int d\omega [g_\gamma(\omega)q^\dagger b_\gamma(\omega) - g_\gamma^\ast(\omega)b_\gamma^\dagger(\omega) q] \\
    &\quad +\int d\omega f_\kappa(\omega) e^{i\omega t}a^\dagger a + \int d\omega f_\gamma(\omega) e^{i\omega t}q^\dagger q.
\end{split}
\end{equation}
The first line describes the exchange interaction between the cavity mode and its bath with operators $b_\kappa(\omega)$ and strength $g_\kappa(\omega)$; the second line accounts for similar interaction between the transmon and its bath. The final line describes dephasing of both modes due to random noise characterized in frequency space by $f_{\kappa,\gamma}(\omega)$. The bath operators have the free Hamiltonian $H_b = \int d\omega\ \omega [b_\kappa^\dagger(\omega)b_\kappa(\omega) + b_\gamma^\dagger(\omega)b_\gamma(\omega)]$.

Transforming the cavity-bath Hamiltonian $H_\kappa = i\int d\omega [g_\kappa(\omega)a^\dagger b_\kappa(\omega) - g_\kappa^\ast(\omega)b_\kappa^\dagger(\omega) a]$ to the dispersive frame gives
\begin{equation}
\begin{split}
    \bar{H}_\kappa &= UH_\kappa U^\dagger \\
    &= i\int d\omega \left[g_\kappa(\omega)\left(a^\dagger + \frac{g}{\Delta}q^\dagger\right)b_\kappa(\omega) -{\rm H.c.}\right].
\end{split}
\end{equation}
Moving to the frame rotating with $H_0 = \omega_a a^\dagger a + \omega_q q^\dagger q + \int d\omega\ \omega b_\kappa^\dagger(\omega)b_\kappa(\omega)$, we can see that the cavity and transmon operators interact with bath operators centered around $\omega_a$ and $\omega_q$, respectively,
\begin{equation}
\begin{split}
    \bar{H}_\kappa &= i\int d\omega g_\kappa(\omega) a^\dagger e^{-i(\omega-\omega_a)t}b_\kappa(\omega) \\
    &\quad +i\int d\omega g_\kappa(\omega) \frac{g}{\Delta}q^\dagger e^{-i(\omega-\omega_q)t}b_\kappa(\omega) + {\rm H.c.}\\
\end{split}
\end{equation}
Following Ref.~\cite{Boissonneault2009}, we now assume that the coupling rates $g_\kappa(\omega)$ do not vary within a sufficiently large bandwidth around $\omega_{a,q}$, although they might differ between these two frequency bands. We can therefore treat the bath operators interacting with the cavity and transmon as independent and the strength of coupling as different for both subsystems while still invoking the Markov approximation. Using standard techniques of quantum optics, we obtain the dissipators
\begin{equation}\label{eq:cav_decay}
\begin{split}
    \mathcal{L}_\kappa\rho &= \kappa(\omega_a)[\bar{n}_\kappa(\omega_a)+1]\mathcal{D}[a]\rho + \kappa(\omega_a)\bar{n}_\kappa(\omega_a)\mathcal{D}[a^\dagger]\rho \\
    &\quad + \kappa(\omega_q)\left(\frac{g}{\Delta}\right)^2[\bar{n}_\kappa(\omega_q)+1]\mathcal{D}[q]\rho \\
    &\quad + \kappa(\omega_q)\left(\frac{g}{\Delta}\right)^2\bar{n}_\kappa(\omega_q)\mathcal{D}[q^\dagger]\rho,
\end{split}
\end{equation}
where $\kappa(\omega) = 2\pi|g_\kappa(\omega)|^2$ and $\bar{n}_\kappa(\omega) = [\exp(\hbar\omega)/(k_BT)-1]^{-1}$ is the thermal occupation of the cavity bath at frequency $\omega$. The first line in Eq.~\eqref{eq:cav_decay} describes the usual cavity damping and heating and the second and third line give the corresponding Purcell decay and heating of the transmon.

Analogous derivation can be done for the transmon bath, $H_\gamma = i\int d\omega [g_\gamma(\omega)q^\dagger b_\gamma(\omega) - g_\gamma^\ast(\omega)b_\gamma^\dagger(\omega) q]$, which results in the Lindblad terms
\begin{equation}
\begin{split}
    \mathcal{L}_\gamma\rho &= \gamma(\omega_q)[\bar{n}_\gamma(\omega_q)+1]\mathcal{D}[q]\rho + \gamma(\omega_q)\bar{n}_\gamma(\omega_q)\mathcal{D}[q^\dagger]\rho \\
    &\quad + \gamma(\omega_a)[\bar{n}_\gamma(\omega_a)+1]\mathcal{D}[a]\rho \\
    &\quad + \gamma(\omega_a)\bar{n}_\gamma(\omega_a)\mathcal{D}[a^\dagger]\rho
\end{split}
\end{equation}
with $\gamma(\omega) = 2\pi|g_\gamma(\omega)|^2$ and $\bar{n}_\gamma(\omega)$ characterizing the thermal noise of the transmon bath at frequency $\omega$. We again get the usual transmon decay and heating and corresponding inverse Purcell terms for the cavity mode.

To evaluate dressed dephasing, we transform the cavity dephasing Hamiltonian $H_{\phi,a} = \int d\omega f_\kappa(\omega)e^{i\omega t}a^\dagger a$,
\begin{equation}
\begin{split}
    \bar{H}_{\phi,a} &= UH_{\phi,a}U^\dagger \\
    &= \int d\omega f_\kappa(\omega)e^{i\omega t}\left[a^\dagger a  + \left(\frac{g}{\Delta}\right)^2q^\dagger q\right] \\
    &\quad +\int d\omega f_\kappa(\omega)\frac{g}{\Delta}e^{i\omega t}(a^\dagger q e^{i\Delta t} + q^\dagger a e^{-i\Delta t}),
\end{split}
\end{equation}
where we also moved to the rotating frame with respect to $H_0$. We can again assume flat (and independent) response within a sufficiently large bandwidth around $\omega = 0,\pm\Delta$ which gives us three independent Lindblad terms,
\begin{equation}
\begin{split}
    \mathcal{L}_{\phi,a}\rho &= \kappa_\phi(\omega\to 0)\mathcal{D}\left[a^\dagger a+\left(\frac{g}{\Delta}\right)^2q^\dagger q\right]\rho \\
    &\quad + \kappa_\phi(\Delta)\left(\frac{g}{\Delta}\right)^2\mathcal{D}[a^\dagger q]\rho + \kappa_\phi(-\Delta)\left(\frac{g}{\Delta}\right)^2\mathcal[q^\dagger a]\rho,
\end{split}
\end{equation}
where $\kappa_\phi(\omega) = 2|f_\kappa(\omega)|^2$~\cite{Boissonneault2009}. From the transmon dephasing Hamiltonian $H_{\phi,q} = \int d\omega f_\gamma(\omega)e^{i\omega t}q^\dagger q$, we get in full analogy
\begin{equation}
\begin{split}
    \mathcal{L}_{\phi,q} &= \gamma_{\phi}(\omega\to 0)\mathcal{D}\left[q^\dagger q+\left(\frac{g}{\Delta}\right)^2a^\dagger a\right]\rho \\
    &\quad + \gamma_\phi(\Delta)\left(\frac{g}{\Delta}\right)^2\mathcal{D}[a^\dagger q]\rho + \gamma_\phi(-\Delta)\left(\frac{g}{\Delta}\right)^2\mathcal{D}[q^\dagger a]\rho.
\end{split}
\end{equation}

Throughout this paper, we assume that the transmon has much stronger dissipation, $\gamma\gg\kappa, \gamma_\phi\gg\kappa_\phi$. We can therefore generally neglect cavity-induced dissipation processes on the transmon and obtain the full master equation
\begin{equation}\label{eq:master_app}
\begin{split}
    \dot{\rho} &= -i[H_0,\rho] + \kappa(\bar{n}_a+1)\mathcal{D}[a]\rho +\kappa\bar{n}_a\mathcal{D}[a^\dagger]\rho\\
    &\quad  + \gamma\left(\frac{g}{\Delta}\right)^2(\bar{n}_q+1)\mathcal{D}[a]\rho +\gamma\left(\frac{g}{\Delta}\right)^2\bar{n}_q\mathcal{D}[a^\dagger]\rho \\
    &\quad   +\gamma(\bar{n}_q+1)\mathcal{D}[q]\rho + \gamma\bar{n}_q\mathcal{D}[q^\dagger]\rho\\
    &\quad   +\gamma_\phi(\Delta)\left(\frac{g}{\Delta}\right)^2\mathcal{D}[a^\dagger q]\rho + \gamma_\phi(-\Delta)\left(\frac{g}{\Delta}\right)^2\mathcal{D}[q^\dagger a]\rho\\
&\quad +\gamma_\phi(\omega\to 0)\mathcal{D}\left[q^\dagger q+\left(\frac{g}{\Delta}\right)^2a^\dagger a\right]\rho+\kappa_\phi\mathcal{D}[a^\dagger a]\rho,
\end{split}
\end{equation}
where we neglect the frequency dependence of $\gamma,\kappa,\bar{n}_{a,q}$ and write $\kappa_\phi = \kappa_\phi(\omega\to 0)$. The correlated dephasing term can be expressed as 
\begin{equation}
\begin{split}
    \mathcal{D}\left[q^\dagger q+\left(\frac{g}{\Delta}\right)^2a^\dagger a\right]\rho = \mathcal{D}[q^\dagger q]\rho + \left(\frac{g}{\Delta}\right)^4\mathcal{D}[a^\dagger a]\rho \\
    \quad + \left(\frac{g}{\Delta}\right)^2\mathcal{S}[a^\dagger a,q^\dagger q]\rho + \left(\frac{g}{\Delta}\right)^2\mathcal{S}[q^\dagger q,a^\dagger a]\rho,
\end{split}
\end{equation}
where $\mathcal{S}[X,Y]\rho = X\rho Y^\dagger - \frac{1}{2}(Y^\dagger X\rho + \rho Y^\dagger X)$.

\section{Master equations for simulations}\label{sec:MEqs}

In this section, we derive the master equations used for numerical simulations of the various control schemes used in the main text. The master equation derived in Appendix~\ref{sec:masterEq} only describes idling; here, we derive additional terms that appear in the Hamiltonian and Lindblad terms due to the presence of different drive tones.

\subsection{Sideband driving}\label{ssec:sideband_app}

We begin with the same Hamiltonian as in Appendix~\ref{sec:mixing} but with only one cavity mode. Following the same sequence of transformations, we end up with
\begin{equation}\label{eq:H0SB}
\begin{split}
H_0 &= (\omega_q -\omega_q')q^\dagger q -\frac{K}{2} q^{\dagger 2} q^2 -K_a a^{\dagger2} a^2  \\
&\quad -\chi_a a^\dagger a q^\dagger q -\chi' a^{\dagger 2} a^2 q^\dagger q \,,
\end{split}
\end{equation}
where $\omega_q'$ is the frequency of the transmon rotating frame. We start with the transmon in logical qubit states $|g\rangle$ and $|e\rangle$. To change the logical space to $\ket{g},\ket{f}$, we introduce a transmon drive at the frequency of the $e-f$ transition ($\omega_1 = \omega_q -K$). By setting the rotating-frame frequency $\omega_q' = \omega_1 = \omega_q-K$, we obtain the total Hamiltonian
\begin{equation}
\begin{split}
H &= K q^\dagger q -\frac{K}{2} q^{\dagger 2} q^2 -K_a a^{\dagger2} a^2  \\
&\quad -\chi_a a^\dagger a q^\dagger q -\chi' a^{\dagger 2} a^2 q^\dagger q +\epsilon_1 q^\dagger  +\epsilon_1^* q.
\end{split}
\end{equation}
The decoherence during this transmon rotation is given by Eq.~\eqref{eq:master_app}.

Next, we drive the transition $|f,0\rangle \to |g,1\rangle$ to transfer the qubit state from the $\{\ket{g},\ket{f}\}$ manifold of the transmon to the cavity, $(\alpha\ket{g}+\beta\ket{f})\ket{0}\to\ket{g}(\alpha\ket{0}+\beta\ket{1})$. We drive the transmon at the frequency $\omega_1 = 2\omega_q -K -\omega_a$, where $2\omega_q -K$ is the frequency of the $g$-$f$ transition. This drives the sideband transition via four-wave mixing
\begin{equation}
H_{\rm int} = -g_{\rm sb}^* a^\dagger q^2 -g_{\rm sb} a q^{\dagger 2},
\end{equation}
where $g_{\rm sb} = \frac{1}{2}K\xi_1g/\Delta$ and we set the rotating frequency $\omega_q' = \omega_q- K/2$. The total Hamiltonian then becomes
\begin{equation}\label{eq:Hsideband}
\begin{split}
H &= \frac{K}{2} (q^\dagger q -q^{\dagger 2} q^2) -K_a a^{\dagger2} a^2 -\chi_a a^\dagger a q^\dagger q \\
&\quad -\chi' a^{\dagger 2} a^2 q^\dagger q -g_{\rm sb}^* a^\dagger q^2 -g_{\rm sb} a q^{\dagger 2}.
\end{split}
\end{equation}

To treat dressed dissipation, we can proceed in direct analogy with Appendix~\ref{sec:masterEq} with one important addition---the pump, which results in displacement $\xi_1 = \epsilon_1/(\omega_1-\omega_q)$ of the transmon mode, serves as an additional dissipation channel. A straightforward calculation then gives the master equation (where we only keep the leading terms) 
\begin{equation}\label{eq:meq_sideband}
\begin{split}
\dot{\rho} &= -i[H,\rho] +\kappa(\bar{n}_a+1)\mathcal{D}[a]\rho + \kappa\bar{n}_a\mathcal{D}[a^\dagger]\rho\\
&\quad +\gamma\left(\frac{g}{\Delta}\right)^2(\bar{n}_q+1)\mathcal{D}[a]\rho +\gamma\left(\frac{g}{\Delta}\right)^2\bar{n}_q\mathcal{D}[a^\dagger]\rho \\
&\quad +\gamma_\phi|\xi_1|^2\left(\frac{g}{\Delta}\right)^2(\mathcal{D}[a]+\mathcal{D}[a^\dagger])\rho \\
&\quad +\gamma(1+\bar{n}_q)\mathcal{D}[q]\rho + \gamma\bar{n}_q\mathcal{D}[q^\dagger]\rho \\
&\quad +\gamma_\phi|\xi_1|^2(\mathcal{D}[q]+\mathcal{D}[q^\dagger])\rho\\
&\quad +\kappa_\phi\mathcal{D}[a^\dagger a]\rho +\gamma_\phi \mathcal{D}\left[q^\dagger q +\left(\frac{g}{\Delta}\right)^2 a^\dagger a\right]\rho \\
&\quad +\gamma_\phi \left(\frac{g}{\Delta}\right)^2(\mathcal{D}[a^\dagger q]\rho+\mathcal{D}[a q^\dagger]\rho)  \,.
\end{split}
\end{equation}
The effects of the transmon dephasing are analogous to Eq.~\eqref{eq:master_app} with the addition of diffusion-like terms for both transmon and cavity (terms proportional to $|\xi_1|^2$ above). During idling, the transmon remains undriven and the dynamics is given by Eq.~\eqref{eq:meq_sideband} with $H = H_0$ and $\xi_1 = 0$.

\subsection{Resonant beam-splitter with coupler}\label{ssec:BS_app}

With a resonant beam-splitter in a driven SNAIL coupler, we have the Hamiltonian
\begin{equation}\label{eq:HcouplerBS}
H = -\frac{K}{2}q^{\dagger 2} q^2 +g_{\rm BS}(a^{\dagger}q + a q^\dagger)
\end{equation}
with transmon Kerr nonlinearity and beam-splitter interaction from a three-wave mixing process $g_{\rm BS} = 6g_3\xi_1(g_a/\Delta_a)(g_q/\Delta_q)$ (where $g_{a,q}$ are coupling rates between the cavity/transmon and coupler and $\Delta_{a,q}$ the corresponding detunings) as described in Appendix~\ref{sec:mixing}. Note that the cavity mode does not inherit a self-Kerr term (or show cross-Kerr interaction with the other modes) since the SNAIL coupler is operated at the Kerr-free point ($g_4 = 0$) and there is no direct interaction between the cavity and transmon.

Dissipation now has to include the effect of the SNAIL coupler bath on the transmon and cavity mode. Thermal noise in the coupler gives rise to inverse Purcell decay in full analogy to direct coupling between transmon and cavity; for the cavity, this can be characterized by the dissipators
\[
    \kappa_{\gamma,c}(\bar{n}_c+1)\mathcal{D}[a]\rho + \kappa_{\gamma,c}\bar{n}_c\mathcal{D}[a^\dagger]\rho,
\]
with inverse Purcell decay rate $\kappa_{\gamma,c} = \gamma_c(g_a/\Delta_a)^2$ (where $\gamma_c$ is the coupler decay rate) and thermal occupation $\bar{n}_c$. The inverse Purcell decay of the transmon can be neglected compared to the intrinsic transmon dissipation since we assume $\gamma\approx\gamma_c$, $\bar{n}_q\approx\bar{n}_c$.

SNAIL dephasing can be treated in analogy to dressed dephasing in directly coupled transmon and cavity with external pump described in the previous section. A straightforward calculation then gives the total master equation (where we keep only leading-order terms)
\begin{equation}\label{eq:MEresonantcoupler}
\begin{split}
    \dot{\rho} &= -i[H,\rho] + \kappa(\bar{n}_a+1)\mathcal{D}[a]\rho + \kappa\bar{n}_a\mathcal{D}[a^\dagger]\rho \\
    &\quad + (\gamma_c+\gamma_{c,\phi})\left(\frac{g_a}{\Delta_a}\right)^2(\bar{n}_c+1)\mathcal{D}[a]\rho \\
    &\quad + (\gamma_c+\gamma_{c,\phi})\left(\frac{g_a}{\Delta_a}\right)^2\bar{n}_c\mathcal{D}[a^\dagger]\rho \\
    &\quad +\gamma_{c,\phi E}|\xi_1|^2\left(\frac{g_a}{\Delta_a}\right)^2(\mathcal{D}[a]+\mathcal{D}[a^\dagger])\rho \\
    &\quad +\left[\kappa_\phi+\gamma_{c,\phi}\left(\frac{g_a}{\Delta_a}\right)^4\right]\mathcal{D}[a^\dagger a]\rho \\
    &\quad +\gamma(\bar{n}_q+1)\mathcal{D}[q]\rho + \gamma\bar{n}_q\mathcal{D}[q^\dagger]\rho + \gamma_\phi\mathcal{D}[q^\dagger q]\rho\\
    &\quad + \gamma_{c,\phi E}|\xi_1|^2\left(\frac{g_q}{\Delta_q}\right)^2(\mathcal{D}[q]+\mathcal{D}[q^\dagger])\rho\\
    &\quad  + \gamma_{c,\phi E}\left(\frac{g_a}{\Delta_a}\right)^2\left(\frac{g_q}{\Delta_q}\right)^2(\mathcal{D}[a^\dagger q]+\mathcal{D}[q^\dagger a])\rho \\
    &\quad +\gamma_{c,\phi}\left(\frac{g_a}{\Delta_a}\right)^2\left(\frac{g_q}{\Delta_q}\right)^2(\mathcal{S}[a^\dagger a,q^\dagger q] +\mathcal{S}[q^\dagger q,a^\dagger a])\rho.
\end{split}
\end{equation}
The SNAIL-dephasing terms here are analogous to Eq.~\eqref{eq:master_app} with three modifications: First, finite hybridization between the transmon and SNAIL gives rise to additional $(g_q/\Delta_q)^2$ factors in collective dissipation terms. Next, tracing out the SNAIL mode in the dressed dephasing dissipators $\mathcal{D}[a^\dagger c]\rho+\mathcal{D}[c^\dagger a]\rho$ gives an additional thermal-noise channel for the cavity. Finally, in analogy with Eq.~\eqref{eq:meq_sideband}, we get additional diffusion-like terms for both cavity and transmon due to the classical pump.

During idling, the pump on the SNAIL is off, the transmon and cavity are fully decoupled (since we assume that the SNAIL is operated at the Kerr-free point), and the master equation reduces to
\begin{equation}\label{eq:SNAILidle}
\begin{split}
    \dot{\rho} &= i\left[\frac{K}{2}q^{\dagger 2}q^2,\rho\right] + \kappa(\bar{n}_a+1)\mathcal{D}[a]\rho + \kappa\bar{n}_a\mathcal{D}[a^\dagger]\rho \\
    &\quad + (\gamma_c+\gamma_{c,\phi})\left(\frac{g_a}{\Delta_a}\right)^2(\bar{n}_c+1)\mathcal{D}[a]\rho \\
    &\quad + (\gamma_c+\gamma_{c,\phi})\left(\frac{g_a}{\Delta_a}\right)^2\bar{n}_c\mathcal{D}[a^\dagger]\rho \\
    &\quad +\left[\kappa_\phi+\gamma_{c,\phi}\left(\frac{g_a}{\Delta_a}\right)^4\right]\mathcal{D}[a^\dagger a]\rho \\
    &\quad +\gamma(\bar{n}_q+1)\mathcal{D}[q]\rho + \gamma\bar{n}_q\mathcal{D}[q^\dagger]\rho + \gamma_\phi\mathcal{D}[q^\dagger q]\rho.
\end{split}
\end{equation}

\subsection{Cavity cascading}

Cavity cascading combines sideband driving with beam-splitter interaction described above. First, an $e$--$f$ rotation is applied on the transmon to transfer the population of the excited state to the second excited state, followed by a sideband transition to transfer the transmon state to a low-$Q$ buffer cavity $b$ as described by the Hamiltonian \eqref{eq:Hsideband}. Afterwards, the state is transferred from the buffer to an ultra-high-$Q$ cavity $a$ using a resonantly driven beam-splitter interaction in a SNAIL coupler; since the buffer cavity, unlike transmon, does not have Kerr nonlinearity, the full Hamiltonian is given just by the beam-splitter coupling, $H = g_{\rm BS}(a^\dagger b+b^\dagger a)$ in contrast to Eq.~\eqref{eq:HcouplerBS}. For recovering the state from the memory, the sequence is inverted---first, beam-splitter interaction is used to swap the state from the storage cavity $a$ to the buffer cavity $b$ and from there to the transmon using a sideband pulse.

The dissipators of the buffer cavity involve its intrinsic relaxation and dephasing as well as those inherited from both transmon and SNAIL. The buffer cavity thus introduces two additional degrees of freedom---its participation ratios in the transmon and SNAIL modes, which can be optimized based on the analytical error derived in Appendix~\ref{sec:fidelities}. The buffer decoherence due to transmon during the sideband pulse is the same as described in Appendix~\ref{ssec:sideband_app}. The dissipation of both cavities due to the driven SNAIL is the same as in Appendix~\ref{ssec:BS_app} with the buffer being additionally subject to processes due to the undriven transmon as derived in Appendix~\ref{sec:masterEq}. Finally, during idling, the state is kept in the storage cavity where it is subject to intrinsic relaxation and decay as well as SNAIL-induced inverse Purcell decay and dephasing as given in Eq.~\eqref{eq:SNAILidle}.

\section{Evaluation of channel fidelities}\label{sec:fidelities}

Analytical expressions for the various channel fidelities can be obtained following the approach of Abad \emph{et al}.~\cite{Abad2022}. We start with derivation of the idling fidelity as that can be obtained directly from the general channel fidelity derived in Ref.~\cite{Abad2022},
\begin{equation}\label{eq:channelFid}
    \bar{\mathcal{F}}_i \approx 1 + \sum_k \Gamma_kt_i \delta F(L_k),
\end{equation}
where, for a single-qubit gate,
\begin{equation}\label{eq:qubitFid}
    \delta F(L) = -\frac{1}{4}{\rm Tr}(L^\dagger L)+\frac{1}{12}\sum_{j\in\{x,y,z\}}{\rm Tr}(L^\dagger\sigma_j L\sigma_j)
\end{equation}
and $L_k$ are jump operators with general decoherence rates $\Gamma_k$. For cavity relaxation ($L_\downarrow = a$), heating ($L_\uparrow = a^\dagger$), and dephasing ($L_\phi = a^\dagger a$), we obtain
\begin{equation}\label{eq:FidIdling}
    \bar{\mathcal{F}}_i \approx 1 - \frac{1}{3}(\Gamma_\downarrow+\Gamma_\uparrow)t_i - \frac{1}{6}\Gamma_\phi t_i.
\end{equation}

For all setups considered here, relaxation is given by the intrinsic damping and inverse Purcell decay, $\Gamma_\downarrow = \kappa+\gamma(g/\Delta)^2$; this expression assumes a zero-temperature bath which agrees well with simulations for $\bar{n}_{a,q}\ll 1$ and which implies $\Gamma_\uparrow \simeq 0$. The total dephasing includes intrinsic and inverse Purcell dephasing and, for cavity directly coupled to the transmon, also a contribution from transmon heating~\cite{Reagor2016}. We therefore get the total dephasing rate $\Gamma_\phi = \kappa_\phi + \gamma_\phi(g/\Delta)^4+\gamma\bar{n}_q$ for sideband driving and $\Gamma_\phi = \kappa_\phi+\gamma_\phi(g/\Delta)^4$ for SNAIL coupler.

When deriving the swap fidelity for the SNAIL coupler, the results of Ref.~\cite{Abad2022} cannot be directly applied since we need to average over all possible initial states of the transmon qubit but need to include three states in the dynamics, $\ket{g,0},\ket{e,0},\ket{g,1}$. In principle, we could treat also the cavity mode as a qubit and the swap as a two-qubit gate, but such an approach would overestimate the total error. In our case, the cavity always starts in the ground state; averaging over all possible qubit states in the cavity would then introduce additional relaxation and dephasing errors.

To derive the swap fidelity, we follow the approach used in Ref.~\cite{Abad2022} to obtain Eqs.~\eqref{eq:channelFid}, \eqref{eq:qubitFid} with appropriate modifications. We assume that the reduction of fidelity is due to a single quantum jump $L_k$ occurring at some point during the gate. The fidelity can be written as
\begin{equation}\label{eq:fidGeneral}
    \bar{\mathcal{F}} = 1 + \sum_k\Gamma_k\int d\psi \bra{\psi}U^\dagger(0,t_{\rm BS})\rho_k U(0,t_{\rm BS})\ket{\psi},
\end{equation}
where we average over all initial transmon qubit states $\ket{\psi}$, $U(0,t_g)$ gives the ideal unitary from time $t = 0$ to the final time $t_{\rm BS}$, and
\begin{equation}
    \rho_k = \int_0^{t_{\rm BS}}dt\, U(t,t_{\rm BS})\{\mathcal{D}[L_k] [U(0,t)\ket{\psi}\bra{\psi}U^\dagger(0,t)]\} U^\dagger(t,t_{\rm BS})
\end{equation}
is the state obtained with a single quantum jump $L_k$ during the gate. Plugging this expression into Eq.~\eqref{eq:fidGeneral}, we obtain
\begin{subequations}
\begin{align}
    \bar{\mathcal{F}} &= 1 + \sum_k\Gamma_k \delta\tilde{F}(L_k), \\
    \delta\tilde{F}(L) &= \int_0^{t_g}dt\int d\psi [\bra{\psi'}L\ket{\psi'}\bra{\psi'}L^\dagger\ket{\psi'}-\bra{\psi'}L^\dagger L\ket{\psi'}],\label{eq:dF}
\end{align}
\end{subequations}
where $\ket{\psi'} = U(0,t)\ket{\psi}$.

The beam-splitter operation transfers an initial excitation from the transmon to the cavity,
\begin{equation}
    \ket{e,0}\to \cos(g_{\rm BS}t)\ket{e,0}+\sin(g_{\rm BS}t)\ket{g,1}.
\end{equation}
Expressing an arbitrary initial state of the transmon in the polar coordinates,
\begin{equation}
    \ket{\psi}_q = \cos\frac{\theta}{2}\ket{g}+e^{i\phi}\sin\frac{\theta}{2}\ket{e},
\end{equation}
we therefore obtain
\begin{equation}
\begin{split}
    \ket{\psi'} &= \cos\frac{\theta}{2}\ket{g,0}+e^{i\phi}\sin\frac{\theta}{2}\cos(g_{\rm BS}t)\ket{e,0} \\
    &\quad +e^{i\phi}\sin\frac{\theta}{2}\sin(g_{\rm BS}t)\ket{g,1}
\end{split}
\end{equation}
from an initial state $\ket{\psi} = \ket{\psi}_q\ket{0}_a$. We can plug this expression into Eq.~\eqref{eq:dF} and use the measure
\begin{equation}
    \int d\psi = \frac{1}{4\pi}\int_0^\pi \sin\theta\, d\theta \int_0^{2\pi}d\phi
\end{equation}
to calculate $\delta\tilde{F}(L)$. Evaluating the integrals is straightforward but lengthy so we omit technical details and only show the main result. We assume that the fidelity is limited primarily by transmon relaxation and dephasing so the total error is obtained by determining $\delta\tilde{F}(q)$ and $\delta\tilde{F}(q^\dagger q)$; we get
\begin{equation}\label{eq:SwapFidelityCoupler}
    \bar{\mathcal{F}}_{\rm BS} = 1-\frac{1}{6}\gamma t_{\rm BS} - \frac{1}{8}\gamma_\phi t_{\rm BS}.
\end{equation}

The same approach can be used to evaluate channel fidelity for sideband driving, where the whole swap consists of an $e$-$f$ rotation of the transmon followed by the sideband transition $\ket{f,0}\to\ket{g,1}$. Both steps have the same three-state structure as the driven coupler; the only difference is in the action of the jump operators $q, q^\dagger q$ on the intermediate states $\ket{\psi'}$. We therefore get
\begin{subequations}
\begin{align}
    \bar{\mathcal{F}}_{ef} &= 1-\frac{7}{12}\gamma t_{ef} - \frac{11}{24}\gamma_\phi t_{ef}, \\
    \bar{\mathcal{F}}_{\rm sb} &= 1 - \frac{1}{2}(\gamma+\gamma_\phi)t_{\rm sb}
\end{align}
\end{subequations}
for the $e$--$f$ rotation and sideband transition, respectively.

Finally, for cavity cascading, we combine a sideband transition between the transmon and a buffer cavity (including the initial $e$--$f$ rotation of the transmon) and a beam-splitter operation between the buffer and storage cavities. The only difference from the above equations is that for the beam-splitter, the infidelity is limited by the relaxation of the buffer cavity (and not the transmon). Eq.~\eqref{eq:SwapFidelityCoupler} is therefore replaced by
\begin{equation}
\begin{split}
    \bar{\mathcal{F}}_{\rm BS} &= 1-\frac{1}{6}\Bigg[(\gamma +\gamma_\phi)\left(\frac{g_q}{\Delta_q}\right)^2 +(\gamma_c +\gamma_{c\phi})\left(\frac{g_b}{\Delta_b}\right)^2 \\
    &\qquad\qquad +\gamma_E|\xi|\left(\frac{g_b}{\Delta_b} \right)^2\Bigg]t_g.
\end{split}
\end{equation}

\

The participation ratios of the buffer cavity in the transmon, $(g_q/\Delta_q)^2$, and the coupler, $(g_b/\Delta_b)^2$, serve as optimization parameters to obtain the minimum total swap error.
During idling, the inverse Purcell rates in Eq.~\eqref{eq:FidIdling} are replaced by the SNAIL-induced inverse Purcell rates.

\vspace{1.25in}

\bibliography{references}

\end{document}